\pdfoutput=1
\documentclass[%
 reprint,
superscriptaddress,
 amsmath,amssymb,
 aps,
]{revtex4-1}
\usepackage{graphicx}
\usepackage{epstopdf}
\begin{document}


\title{MBE preparation and \textit{in situ} characterization of FeTe thin films}


\author{V.M. Pereira}
\affiliation{Max Planck Institute for Chemical Physics of Solids, N{\"o}thnitzer Str. 40, 01187 Dresden, Germany}

\author{C.N. Wu}
\affiliation{Max Planck Institute for Chemical Physics of Solids, N{\"o}thnitzer Str. 40, 01187 Dresden, Germany}
\affiliation{Department of Physics, National Tsing Hua University, Hsinchu 30013, Taiwan}

\author{C.E. Liu}
\affiliation{Max Planck Institute for Chemical Physics of Solids, N{\"o}thnitzer Str. 40, 01187 Dresden, Germany}
\affiliation{Department of Electrophysics, National Chiao Tung University, Hsinchu 300, Taiwan}

\author{S.-C. Liao}
\affiliation{Max Planck Institute for Chemical Physics of Solids, N{\"o}thnitzer Str. 40, 01187 Dresden, Germany}

\author{C.F. Chang}
\affiliation{Max Planck Institute for Chemical Physics of Solids, N{\"o}thnitzer Str. 40, 01187 Dresden, Germany}

\author{C.-Y. Kuo}
\affiliation{Max Planck Institute for Chemical Physics of Solids, N{\"o}thnitzer Str. 40, 01187 Dresden, Germany}
\affiliation{National Synchrotron Radiation Research Center, Hsinchu 30076, Taiwan}

\author{C. Koz}
\affiliation{Max Planck Institute for Chemical Physics of Solids, N{\"o}thnitzer Str. 40, 01187 Dresden, Germany}

\author{U. Schwarz}
\affiliation{Max Planck Institute for Chemical Physics of Solids, N{\"o}thnitzer Str. 40, 01187 Dresden, Germany}

\author{H.-J. Lin}
\affiliation{National Synchrotron Radiation Research Center, Hsinchu 30076, Taiwan}

\author{C.T. Chen}
\affiliation{National Synchrotron Radiation Research Center, Hsinchu 30076, Taiwan}

\author{L.H. Tjeng}
\affiliation{Max Planck Institute for Chemical Physics of Solids, N{\"o}thnitzer Str. 40, 01187 Dresden, Germany}

\author{S.G. Altendorf}
\affiliation{Max Planck Institute for Chemical Physics of Solids, N{\"o}thnitzer Str. 40, 01187 Dresden, Germany}

\date{\today}

\begin{abstract}
We have synthesized Fe$_{1+y}$Te thin films by means of molecular beam epitaxy (MBE) under
Te-limited growth conditions. We found that epitaxial layer-by-layer growth is possible for a wide 
range of excess Fe values, wider than expected from what is known from studies on the bulk material. 
Using x-ray magnetic circular dichroism spectroscopy at the Fe $L_{2,3}$ and Te $M_{4,5}$ edges, 
we observed that films with high excess Fe contain ferromagnetic clusters while films with lower excess Fe remain nonmagnetic. Moreover, x-ray absorption spectroscopy showed that it is possible to obtain films with very similar electronic structure as that of a high quality bulk single crystal Fe$_{1.14}$Te. 
Our results suggest that MBE with Te-limited growth may provide an opportunity to synthesize 
FeTe films with smaller amounts of excess Fe as to come closer to a possible superconducting phase.
\end{abstract}


\maketitle

\section{Introduction}

Within the family of Fe-based superconductors \cite{Kamihara2008, Zhao2008, Hsu2008}, FeSe has been one of the most studied materials due to its simple binary crystal structure \cite{Hsu2008}. Its sister compound FeTe has received considerable attention as well, but surprisingly the latter is not superconducting. Instead, FeTe is an insulator that orders antiferromagnetically at low temperatures \cite{Bao2009, Li2009}. 
Also astonishing is the contrasting behavior upon application of external pressure: while FeSe enhances its superconducting temperature \cite{Mizuguchi2008, Margadonna2009, Medvedev2009}, FeTe becomes ferromagnetic when it turns into a metal at high pressures \cite{Ciechan2013, Mydeen2017}. 
Adding to the complexity, bulk FeTe only forms in the presence of excess Fe, i.e. Fe$_{1+y}$Te with $ 0.04 \lessapprox  y \lessapprox 0.17$ \cite{Rodriguez2011, Roessler2011, Koz2013}, which is quite different from FeSe, which is known to exhibit a negligible homogeneity range only. 

FeTe in thin film form has also been studied. Many of the investigations were carried out with the focus on how one can induce superconductivity using thin films \cite{Han2010}. It came as a surprise that a partial oxidation of the Fe$_{1+y}$Te thin films by exposure to oxygen can indeed induce superconductivity \cite{Nie2010, Si2010, Telesca2012,Zheng2013}. This has to be contrasted with the observation that the superconductivity in FeSe thin films is destroyed by oxygen exposure \cite{Telesca2012}. For bulk FeTe samples the situation is more complicated. It seems that superconductivity can be induced by reaction with oxygen or other chemicals only if there is also some minimum amount of S or Se in the system \cite{Dong2011, Hu2012, Kawasaki2012}. The reasons behind the mechanism of superconductivity upon oxygen exposure are still not well understood, although there are some indications that the variation of interstitial Fe might play a crucial role  \cite{Hu2012}. 

Pulsed laser deposition (PLD) \cite{Han2010, Nie2010, Si2010, Telesca2012} and molecular beam epitaxy (MBE) \cite{Zheng2013, Hu2014, Li2016} have been applied to synthesize the films. In these earlier PLD studies, nominally stoichiometric targets or targets with some excess Te have been used. In the MBE investigations, the synthesis of the films was effectively carried out under the so-called Fe-limited growth conditions, in which the Te is deposited in excess and, at the same time, most of this excess is re-evaporated by using sufficiently high substrate temperatures.

In this study we explored a different route for the preparation of FeTe thin films. We applied the Te-limited growth in our MBE process. Our objective is to investigate the stability range of the excess Fe for the formation of FeTe films. By determining the electronic and magnetic properties of the Fe$_{1+y}$Te films we aim to gain better insight into the growth properties of the films and to estimate whether films can be made with very small amounts of excess Fe, as to come closer to a possible superconducting phase. The structural, electronic and magnetic characterization measurements were carried out \textit{in situ} in order to ensure reliable data.

\begin{figure*}[htbp]
\minipage{0.25\textwidth}
 \includegraphics[width=\linewidth]{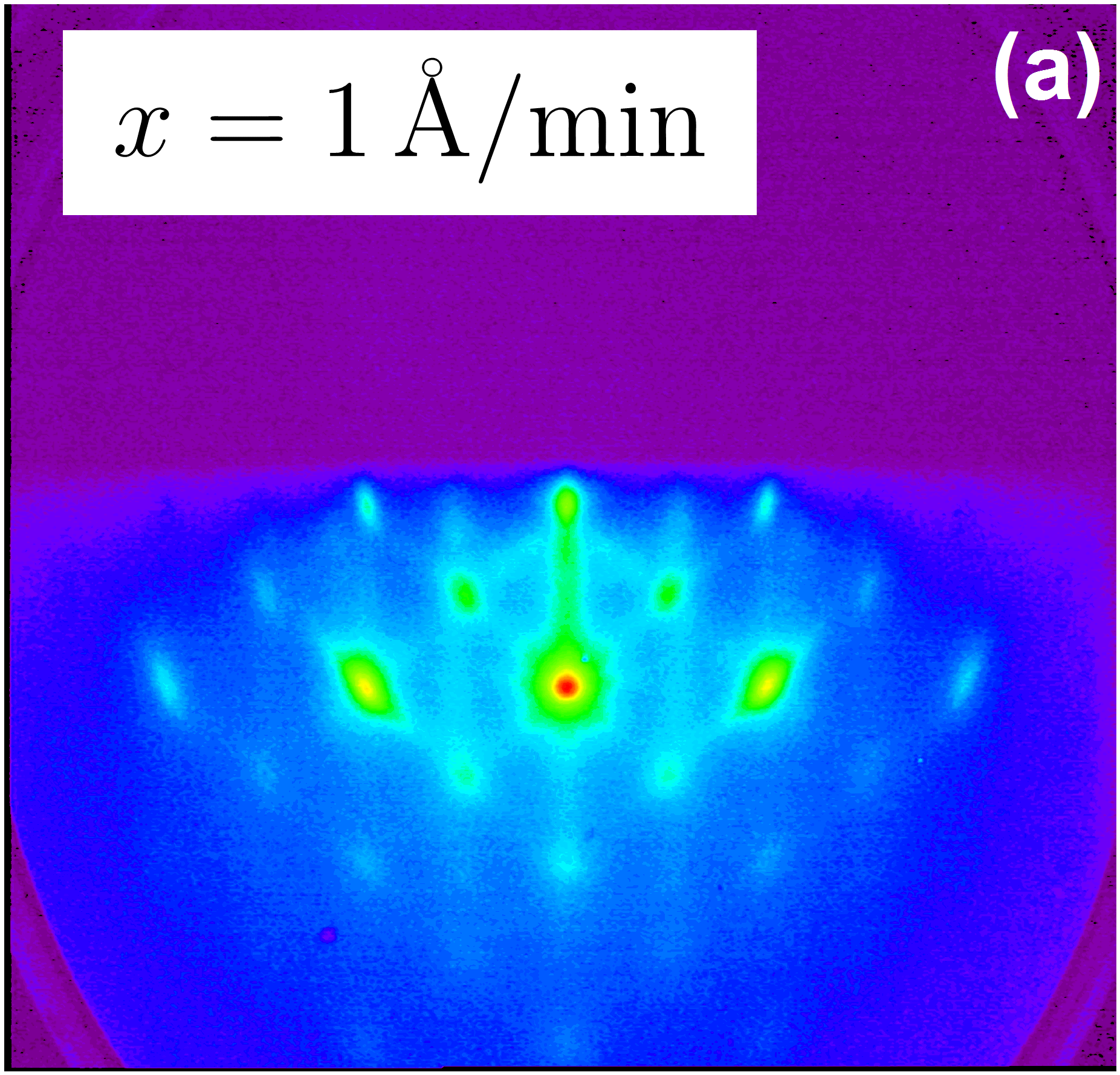}
\endminipage\hfill
\minipage{0.25\textwidth}
  \includegraphics[width=\linewidth]{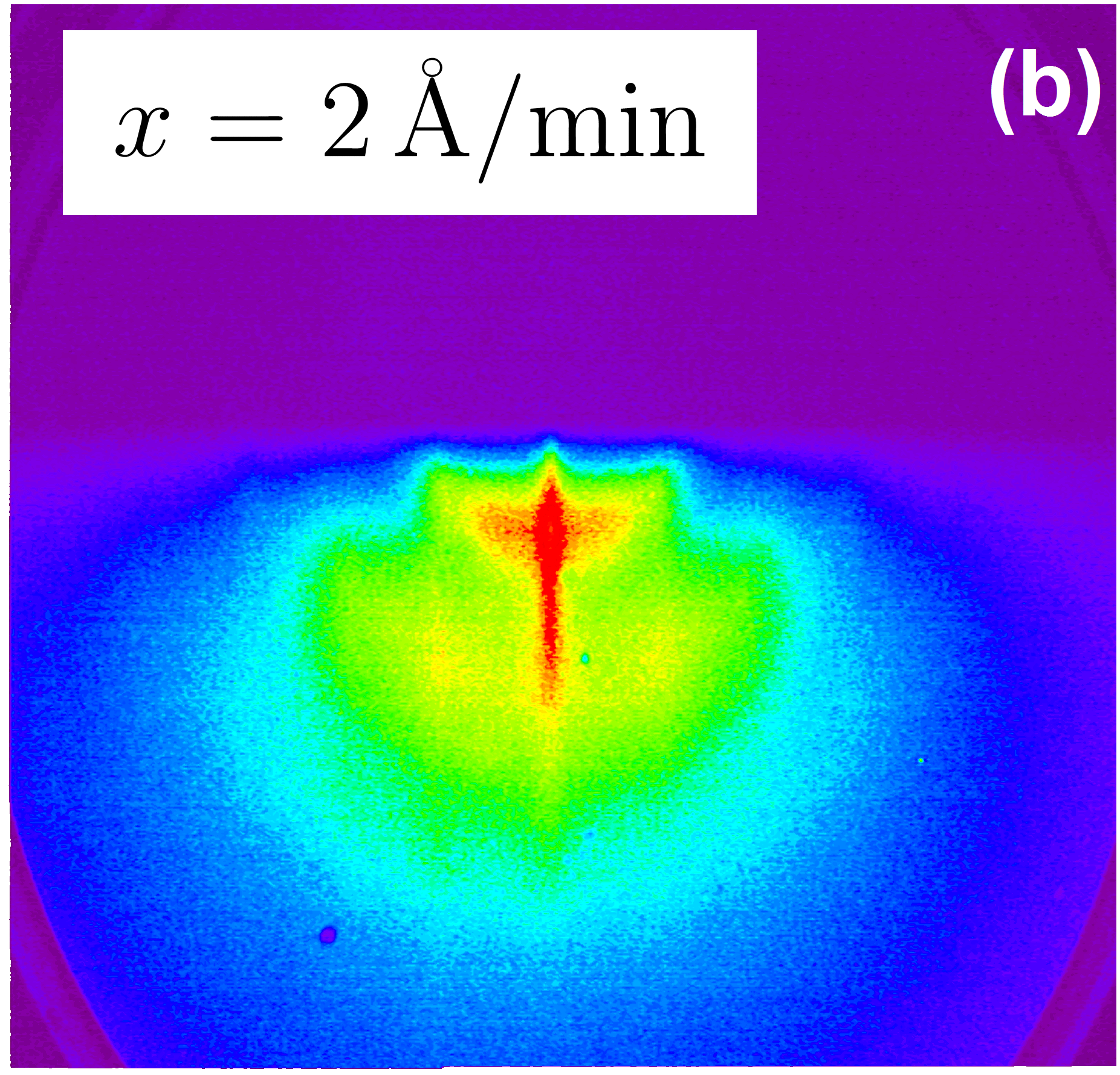}
\endminipage\hfill
\minipage{0.25\textwidth}%
  \includegraphics[width=\linewidth]{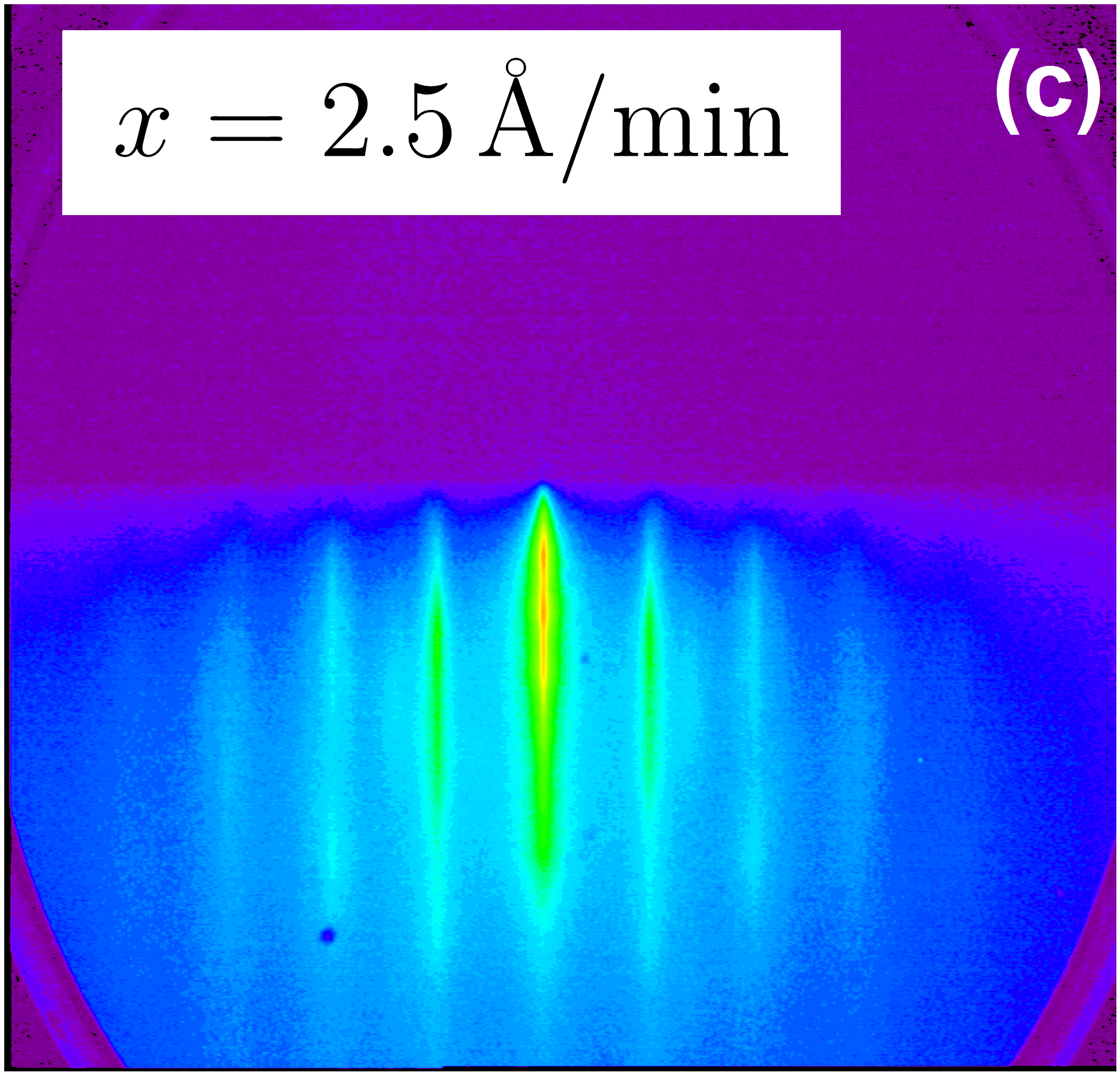}
\endminipage\hfill
\minipage{0.25\textwidth}%
  \includegraphics[width=\linewidth]{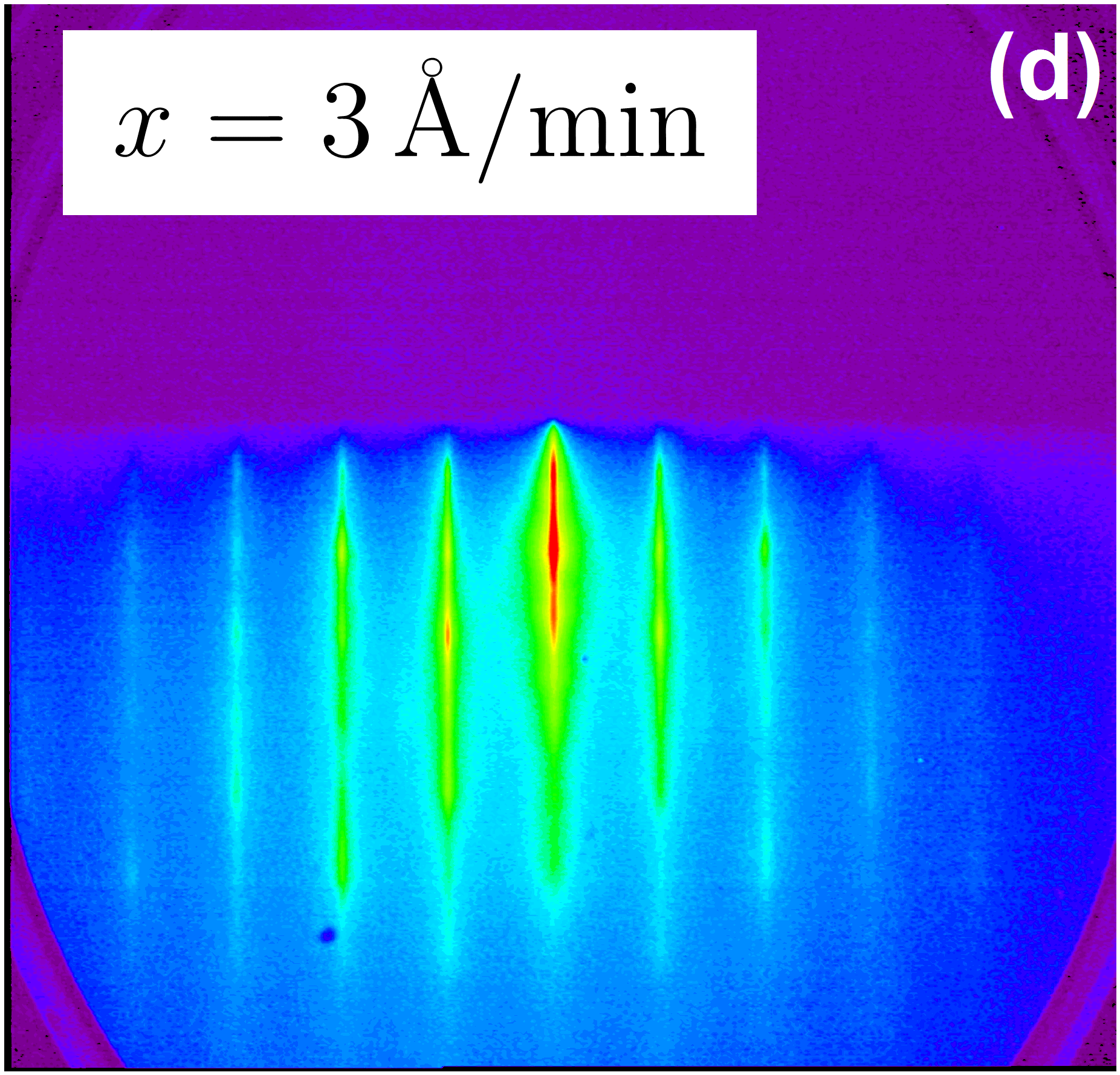}
\endminipage \\
\minipage{0.25\textwidth}%
  \includegraphics[width=\linewidth]{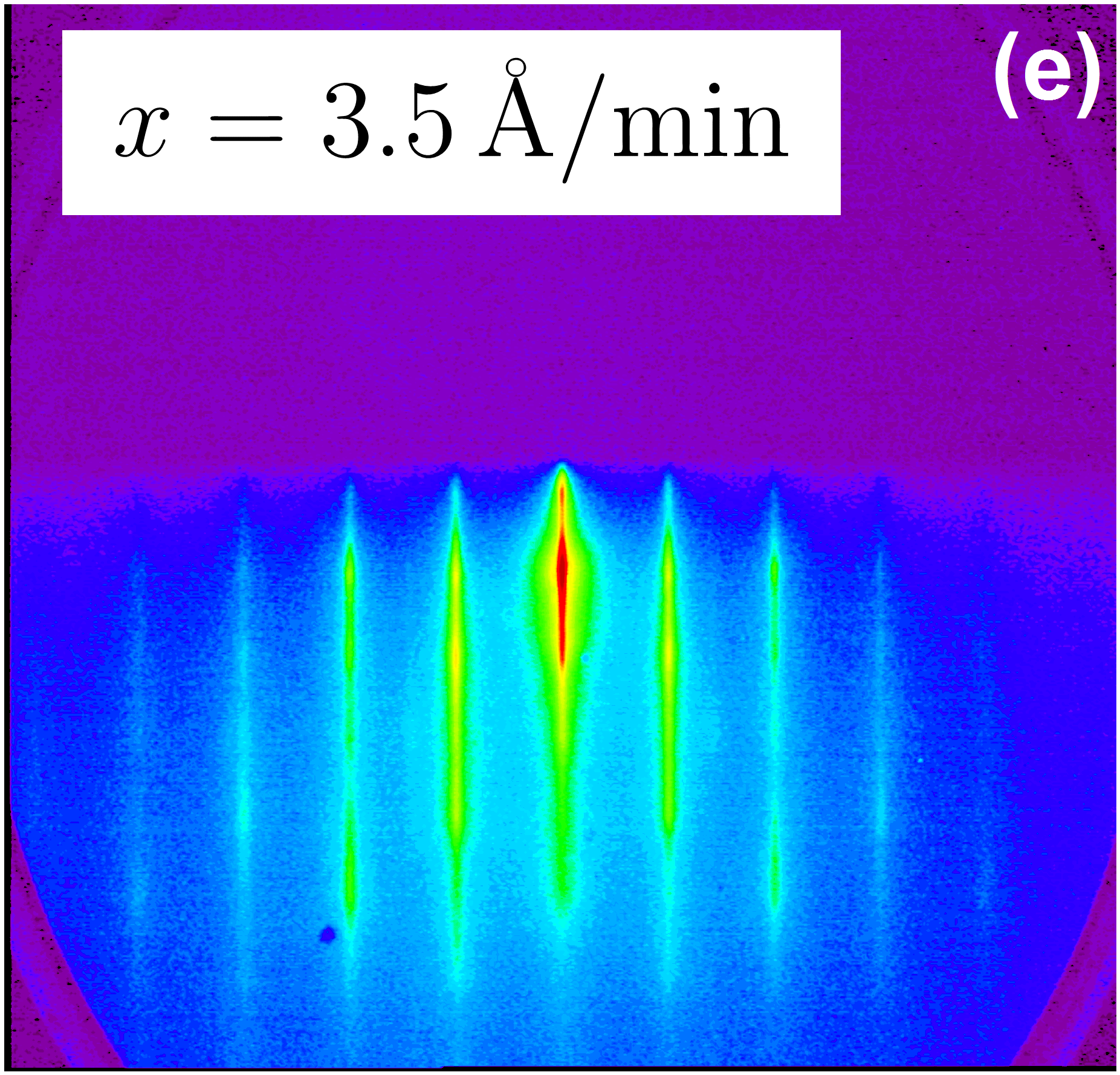}
\endminipage\hfill
\minipage{0.25\textwidth}%
  \includegraphics[width=\linewidth]{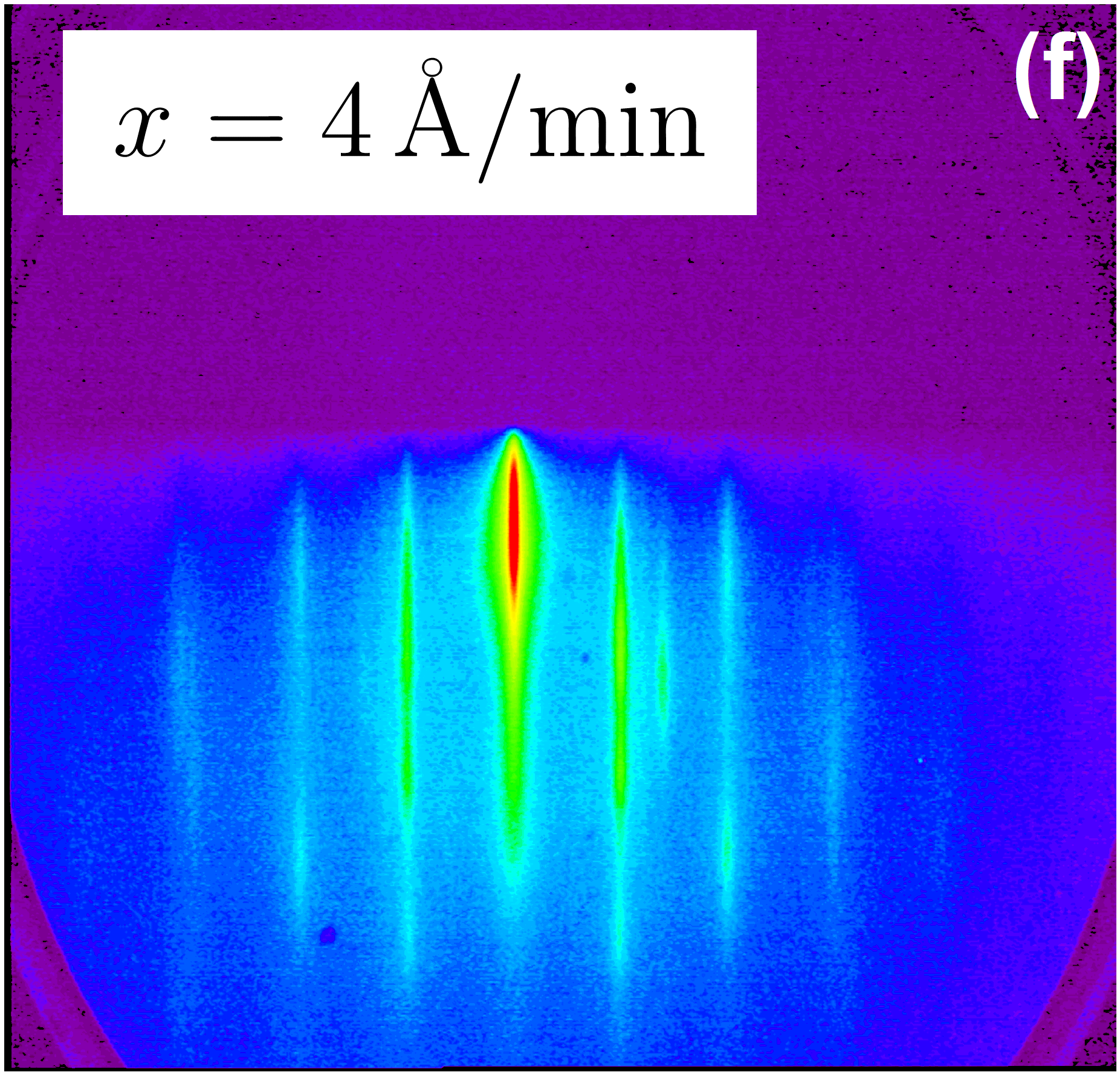}
\endminipage\hfill
\minipage{0.25\textwidth}%
  \includegraphics[width=\linewidth]{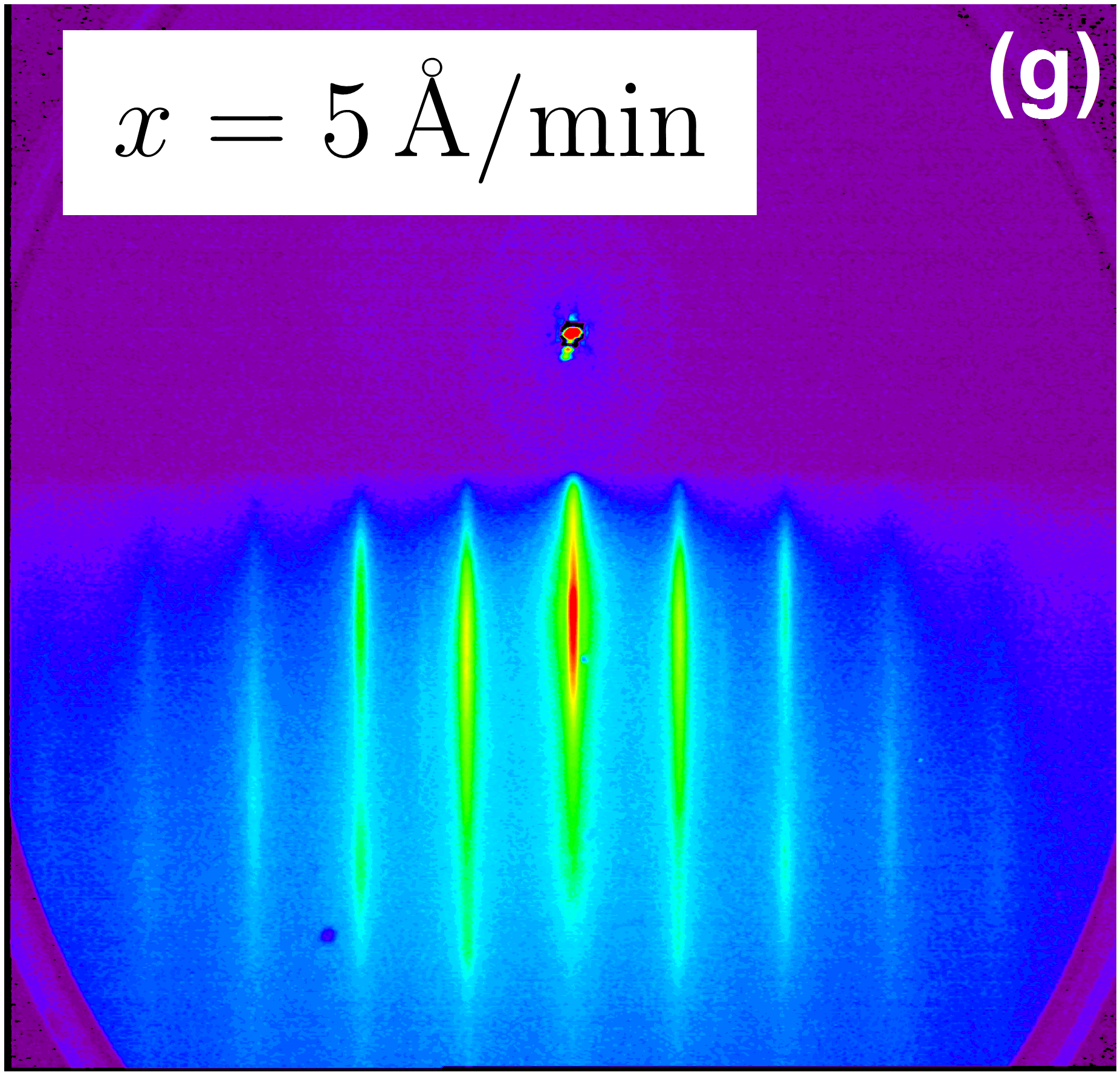}
\endminipage\hfill
\minipage{0.25\textwidth}%
  \includegraphics[width=\linewidth]{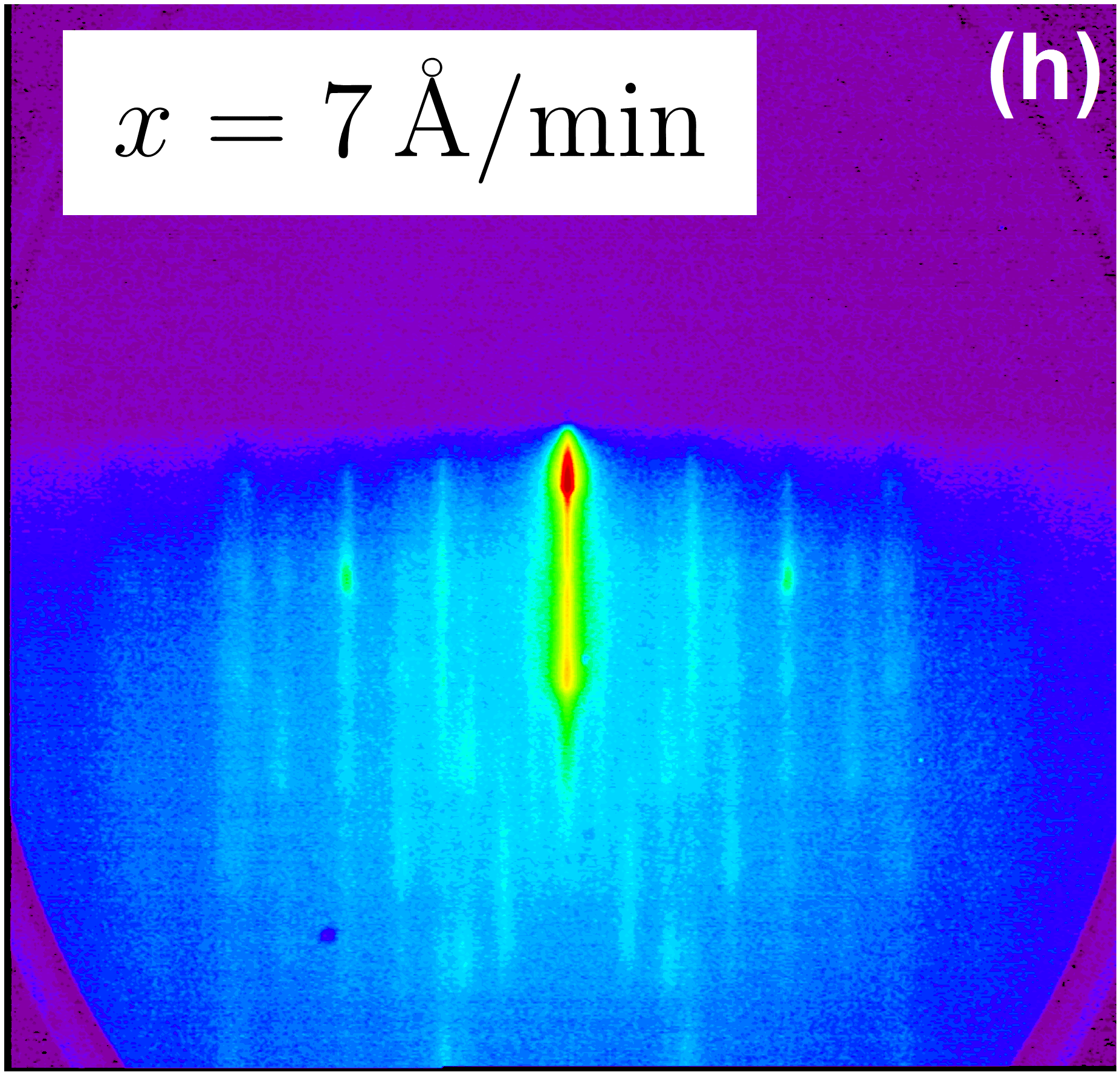}
\endminipage
\caption{RHEED patterns after 30 min of FeTe films growth. The Te deposition rate $x$ varies among (a) 1, (b) 2, (c) 2.5, (d) 3, (e) 3.5, (f) 4, (g) 5, and (h) 7 \AA/min, while the Fe rate is fixed for all films at 1 \AA/min.} \label{fig_RHEED}
\end{figure*}

\section{Experiment}

\normalsize
The FeTe films were grown by MBE on epi-polished  MgO (001) substrates purchased from CrysTec GmbH. Prior to the deposition, the substrates were annealed at 600 $^\circ$C for 2 hours in an oxygen atmosphere of 3$\times$10$^{-7}$ mbar. The substrate temperature was kept at 250 $^\circ$C during growth. 

For all the reported films, the growth time was fixed at 30 minutes. Fe and Te were evaporated from effusion cells. The flux rates of the Fe and Te were measured by a quartz crystal monitor at the growth position. The Fe flux rate was set at 1 \AA/min (T$_{\rm{Fe}}$ $\sim$ 1220 $^\circ$C effusion cell temperature) for all films, while the Te flux rate (named $x$ throughout the report) was varied between 1 to 7 \AA/min (T$_{\rm{Te}}$ $\sim$ 245-265 $^\circ$C effusion cell temperature). The base pressure of the MBE system was 5$\times$10$^{-10}$ mbar. 

\textit{In situ} and \textit{real-time} monitoring of the epitaxial growth was performed by reflection high-energy electron diffraction (RHEED) measurements using a STAIB Instruments RH35 system. The kinetic energy of the electrons was set at 20 keV. X-ray absorption spectroscopy (XAS) and x-ray magnetic circular dichroism (XMCD) measurements at the Fe $L_{2,3}$ and Te $M_{4,5}$ edges were carried out \textit{in situ} at the 11A Dragon beam line of the Taiwan Light Source at the NSRRC in Taiwan. The spectra were recorded at room temperature in the total electron yield mode with a photon-energy resolution of 0.25 eV.  An Fe$_2$O$_3$ and a Cr$_2$O$_3$ crystal 
were measured simultaneously in a separate chamber to obtain relative energy referencing with an accuracy better than a few meV  for the Fe $L_{2,3}$ ($h\nu =700-730$ eV) and Te $M_{4,5}$ ($h\nu =565-590$ eV) spectra. The Fe $L_3$ white line of Fe$_2$O$_3$ is set at 706.2 eV and the first peak of the Cr $L_3$ of Cr$_2$O$_3$ at 574.7 eV. A magnetic field of about 0.28 T was applied. The pressure of the XAS/XMCD chamber was 2$\times$10$^{-10}$ mbar during the measurements. In order to compare the electronic properties of the films, we used a bulk sample with composition Fe$_{1.14}$Te, synthesized by chemical vapor transport as described in \cite{Roesler2016}. The XAS measurements on this sample were performed at the TPS 45A beam line of the Taiwan Photon Source at the NSRRC in Taiwan. 

\textit{Ex situ} x-ray diffraction (XRD) measurements were used to further characterize the structural quality of the films. The measurements were performed with a PANalytical X'Pert PRO diffractometer using monochromatic Cu-$K_{\alpha1}$ radiation ($\lambda$ = 1.54056 \AA).

\section{Results}
Figure \ref{fig_RHEED} shows the RHEED patterns of the films prepared with the Fe deposition rate 
fixed at 1 \AA/min and with Te deposition rates $x$ varied between 1 and 7 \AA/min. These RHEED 
pictures were all taken after 30 min of deposition. One can observe spots and rings for the film 
prepared with $x$ = 1 \AA/min (a), indicating polycrystalline and island growth. For $x$ = 2 \AA/min (b), some faint streaky patterns can be seen on top of a high background, and for 
$x$ = 2.5, 3, 3.5, 4, and 5 \AA/min (c-g) the same streaky patterns become much clearer. 
For $x$ = 7 \AA/min (h), new additional streaks as well as spots start to appear. The results indicate that for $x$ = 1 and 7 \AA/min the films consist of several different phases. For $x$ = 2, 2.5, 3, 3.5, 4 and 5 \AA/min, the appearance of the same streaky patterns is in agreement with the presence of the Fe$_{1+y}$Te phase of the bulk, although the amount and quality could vary depending on how clear and crisp the patterns are with respect to the background. We note that, although MgO (001) has a high lattice mismatch ($\sim$ 9 \%), the layered structure of FeTe relies on Van der Waals epitaxy, which enables a crystalline growth of the films. 

\begin{figure}[htbp]
\minipage{0.7\columnwidth}
	\includegraphics[width=1\textwidth]{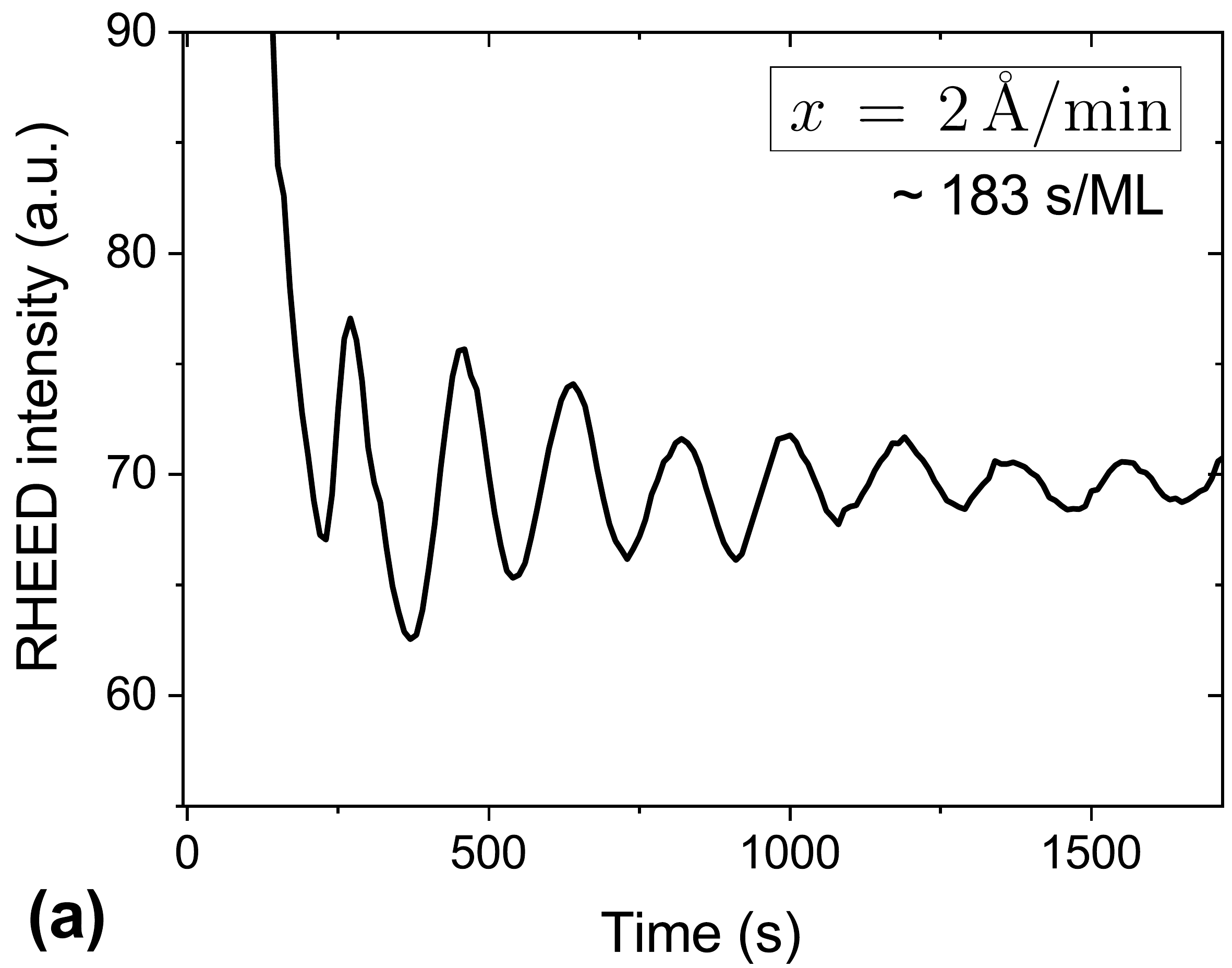}
\endminipage\hfill
\minipage{0.7\columnwidth}
	\includegraphics[width=1\textwidth]{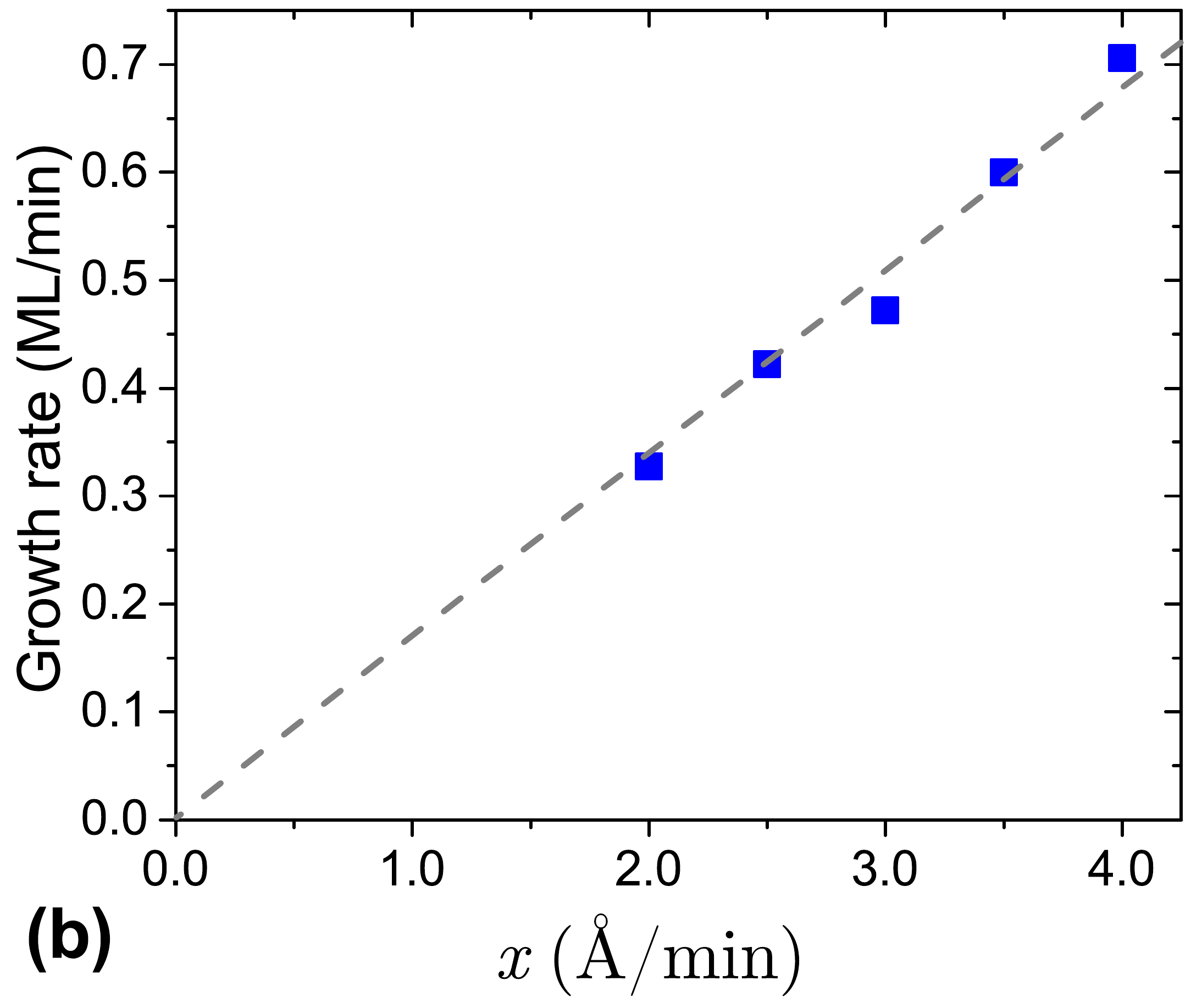}
\endminipage\hfill
\caption{(a) Oscillations of the RHEED specular spot for a film with $x$ = 2 \AA/min.  (b) Dependence of the growth rate of the FeTe thin films with the Te deposition rate $x$ as determined from the time period of the RHEED oscillations. The dashed line is a linear fit with intercept fixed at 0.}\label{fig_RHEED_osc}
\end{figure}

To obtain more information about the growth process, we have monitored the intensity of the specular spot of the RHEED as a function of time. We have been able to observe intensity oscillations for the films prepared with $x$ = 2, 2.5, 3, 3.5 and 4 \AA/min. In Figure \ref{fig_RHEED_osc}(a) we show an example of such oscillations for a film with $x$ = 2 \AA/min. These pronounced oscillations indicate a layer-by-layer epitaxial growth. We found that the oscillation time varies with $x$. It is about 183 seconds per monolayer (s/ML) for the $x$ = 2 \AA/min film and becomes shorter for higher $x$ values. Converting these numbers into monolayer per minute (ML/min), we then can plot the growth rate as a function of $x$. This is displayed in Figure \ref{fig_RHEED_osc}(b). We can notice a strong linear relationship between the growth rate and
the Te deposition rate, establishing that the growth is controlled by the Te supply. We are thus operating in a Te-limited growth mode. This is to be contrasted to the earlier thin film studies where the Fe supply was limiting the growth \cite{Zheng2013, Hu2014, Li2016}.
This is caused by the low substrate temperature relative to the Te supply rate used in our study. 

\begin{figure}[htbp]
\minipage{0.7\columnwidth}
	\includegraphics[width=\textwidth]{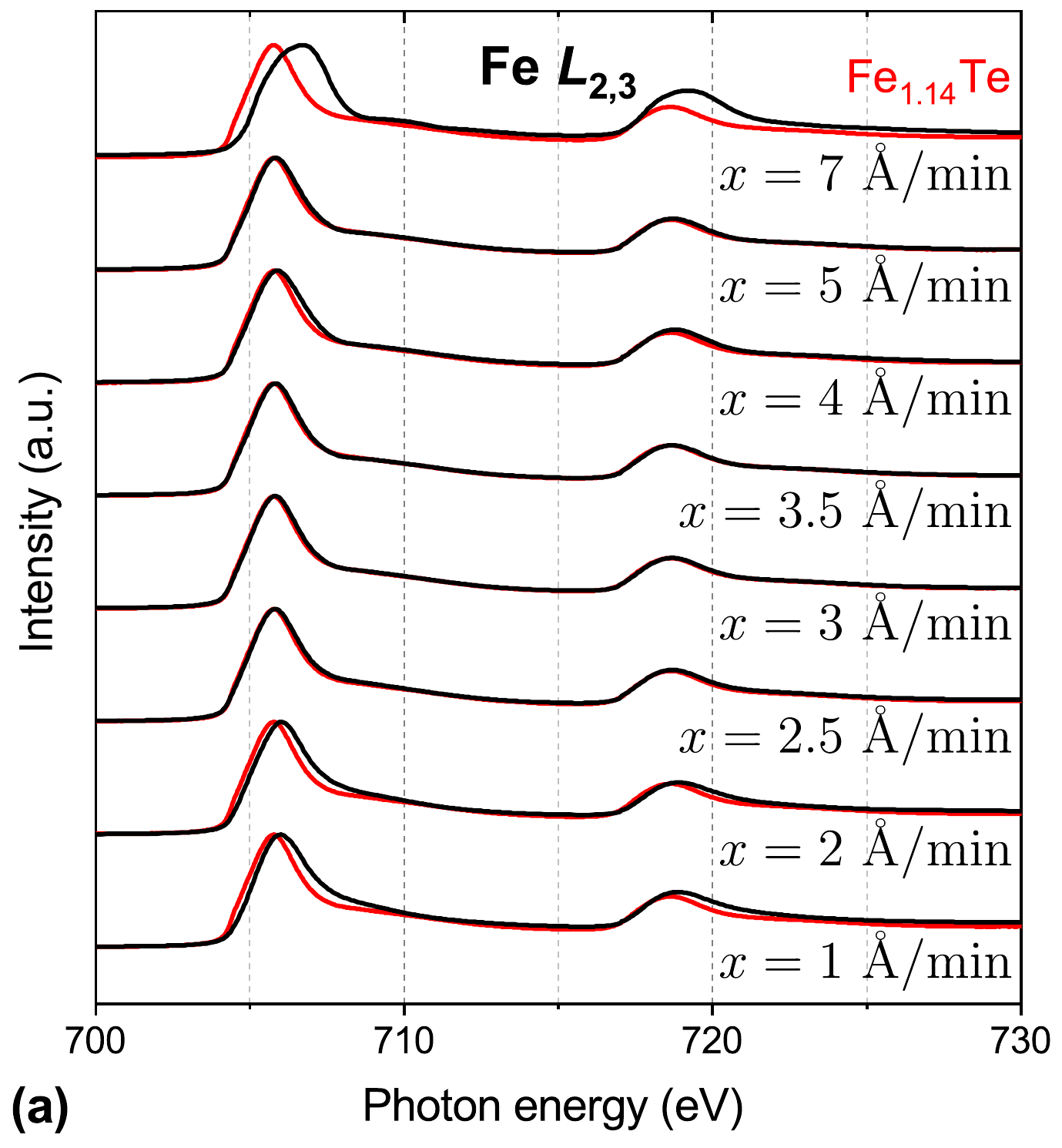}
\endminipage\hfill
\minipage{0.7\columnwidth}
	\includegraphics[width=\textwidth]{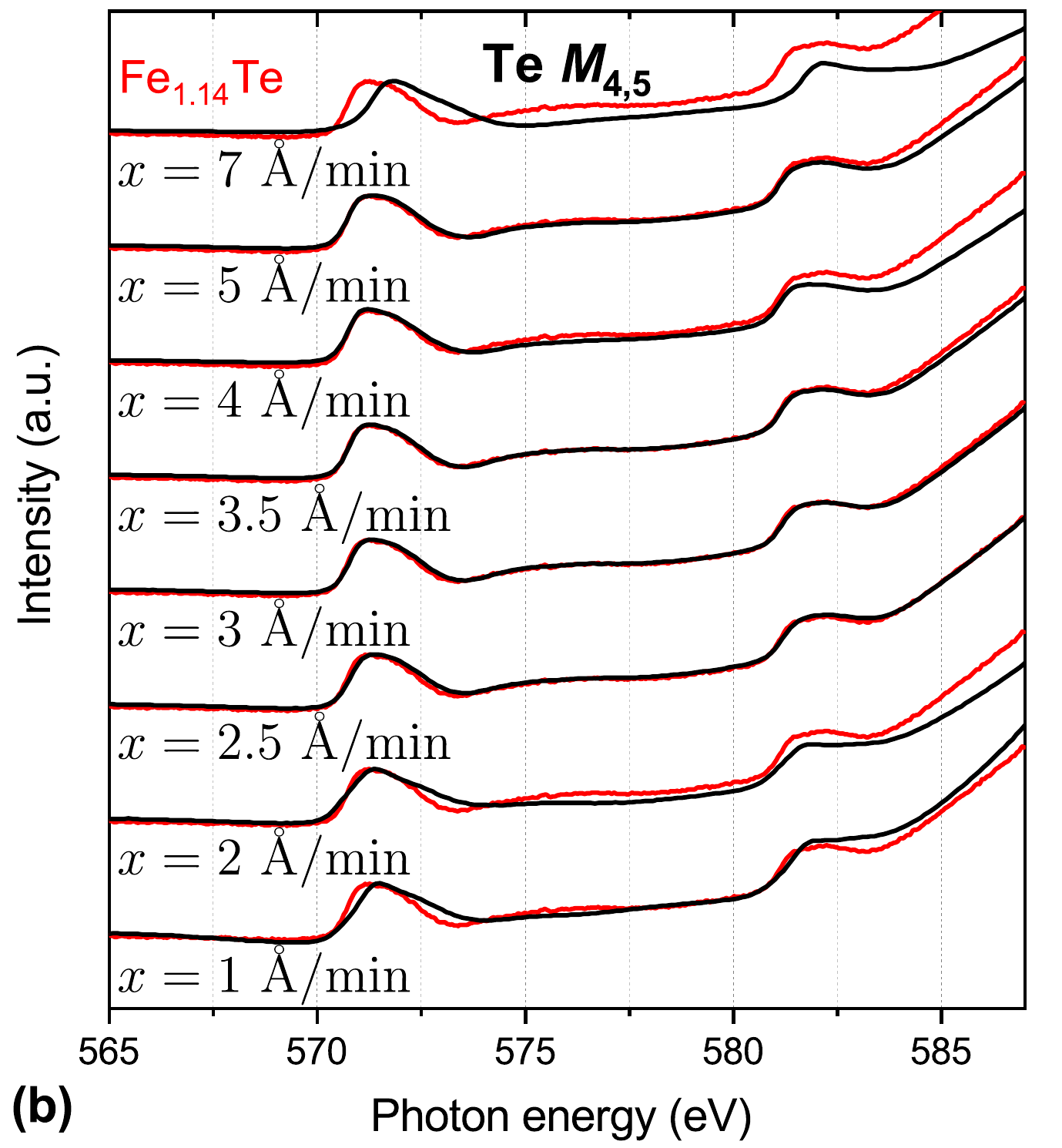}
\endminipage
\caption{(a) Fe $L_{2,3}$ and (b) Te $M_{4,5}$ XAS spectra of the FeTe thin films prepared with varying Te deposition rates $x$ (black) in comparison to the XAS spectra of single crystalline bulk Fe$_{1.14}$Te (red).}\label{fig_XAS}
\end{figure}

\begin{figure*}[htbp]
\minipage{0.32\textwidth}
	\includegraphics[width=\textwidth]{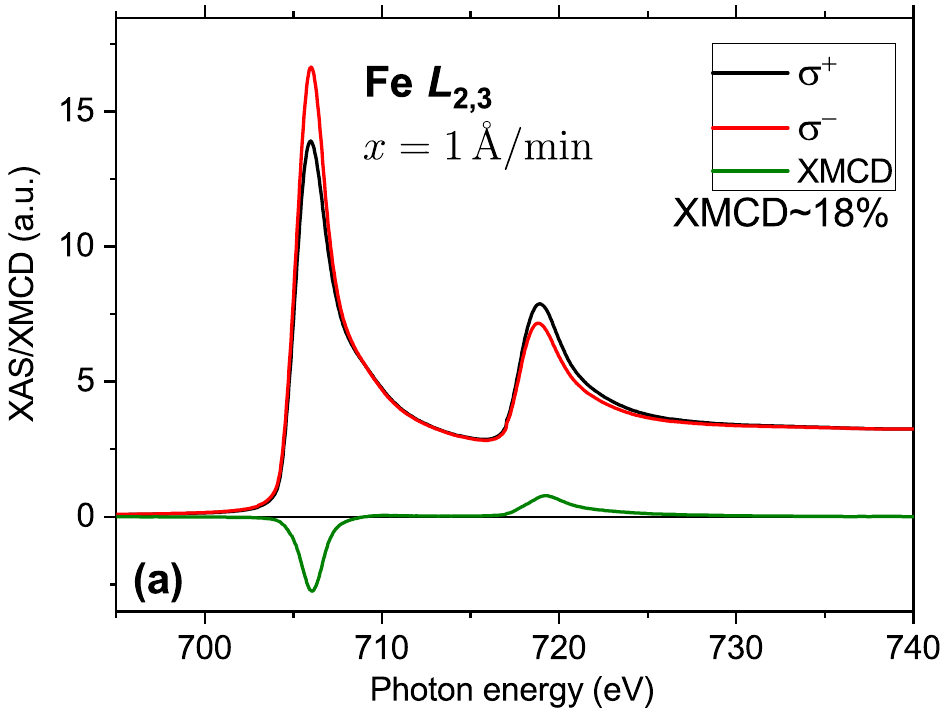}
\endminipage\hfill
\minipage{0.32\textwidth}
	\includegraphics[width=\textwidth]{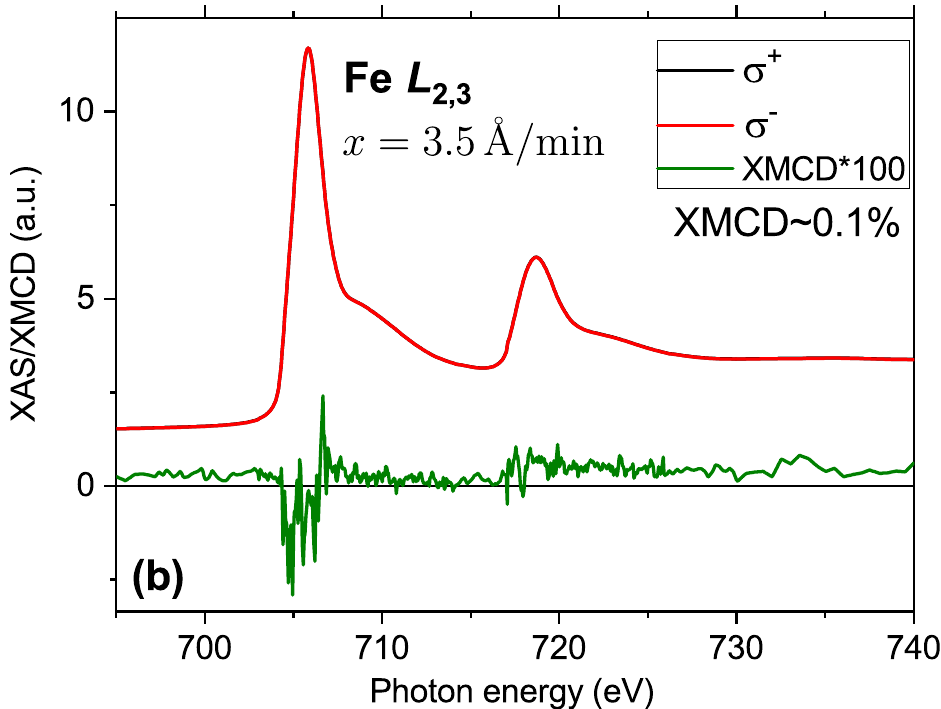}
\endminipage\hfill
\minipage{0.32\textwidth}
	\includegraphics[width=\textwidth]{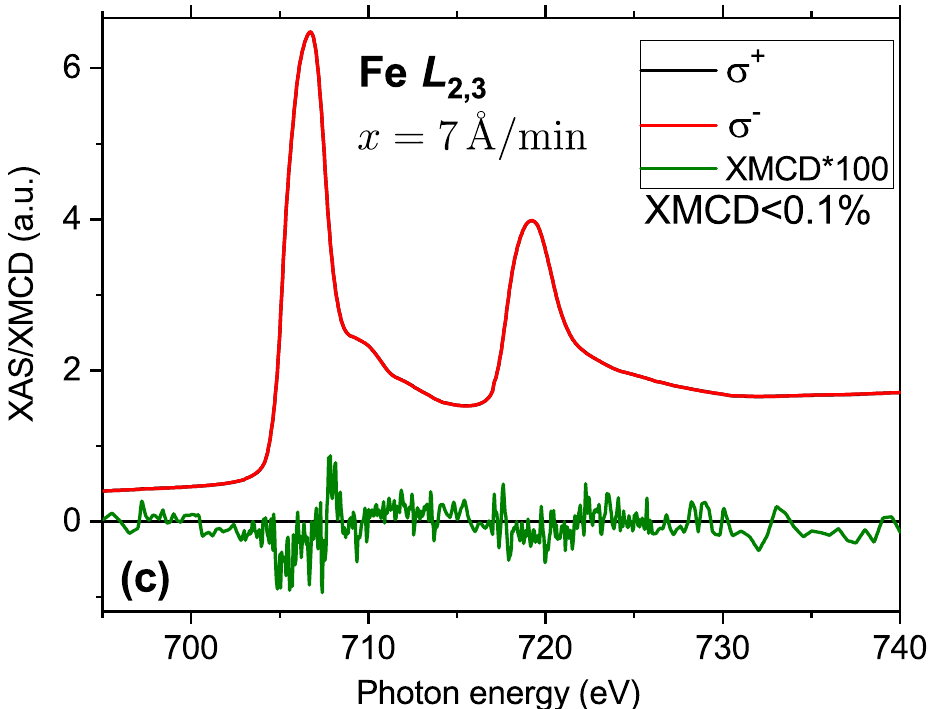}
\endminipage \\
\minipage{0.32\textwidth}
	\includegraphics[width=\textwidth]{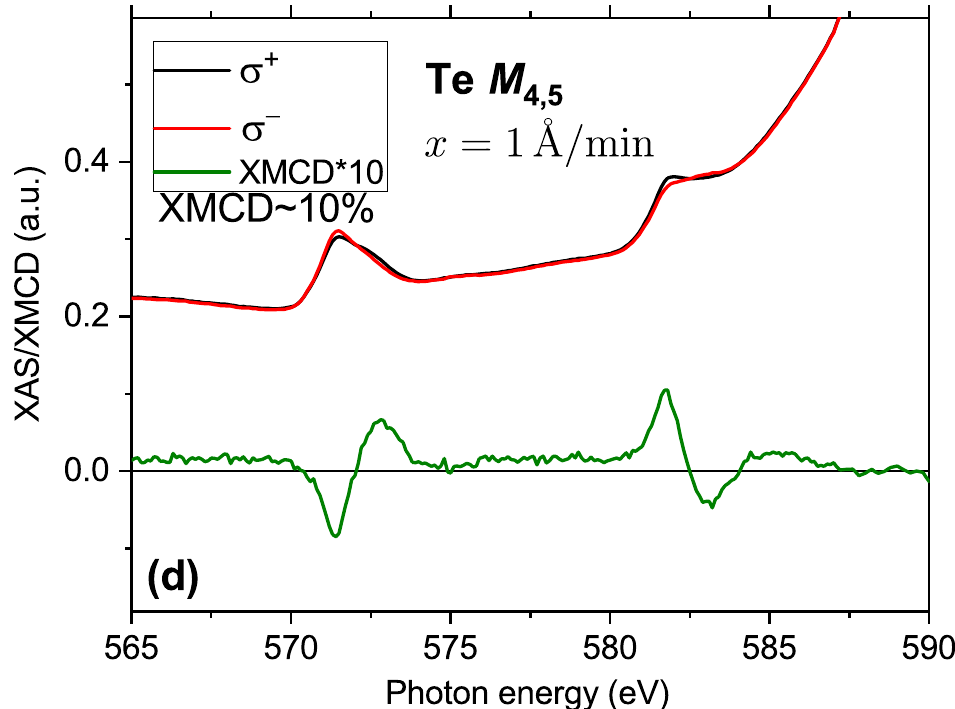}
\endminipage\hfill
\minipage{0.32\textwidth}
	\includegraphics[width=\textwidth]{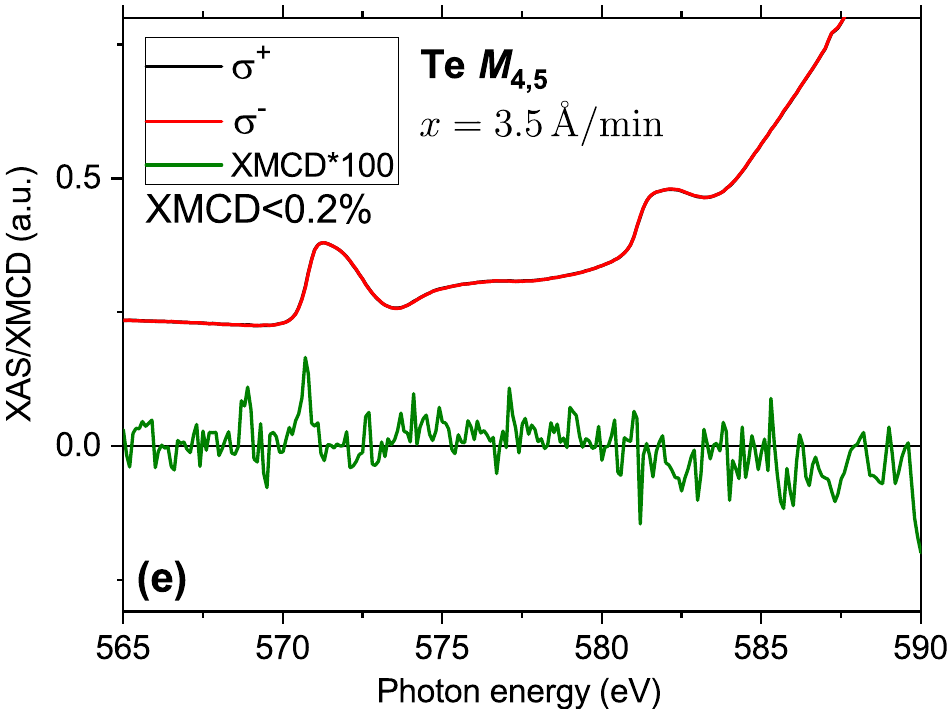}
\endminipage\hfill
\minipage{0.32\textwidth}
	\includegraphics[width=\textwidth]{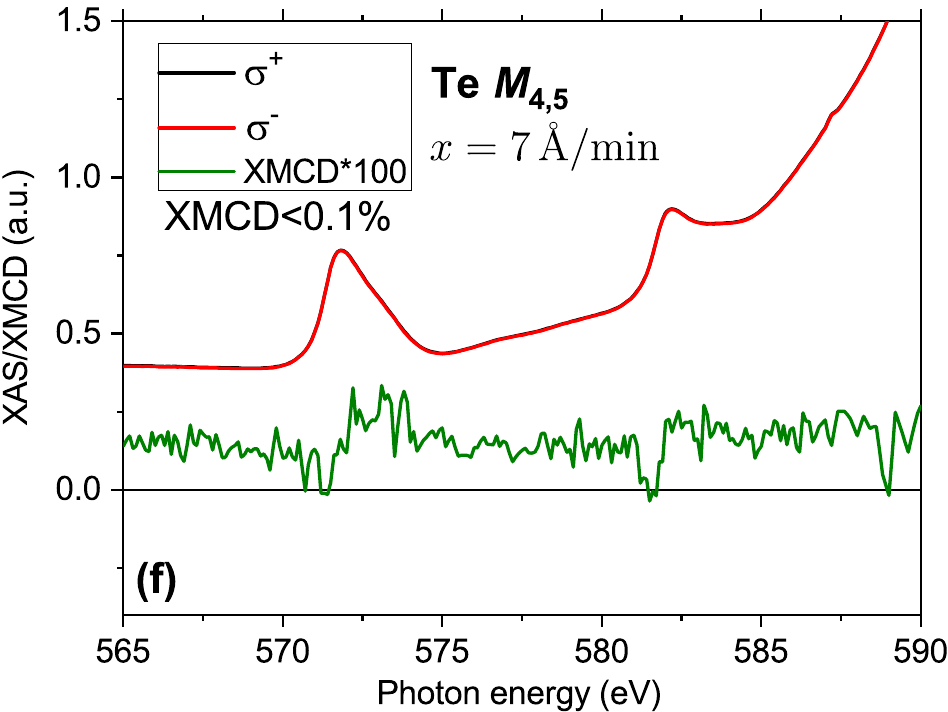}
\endminipage
\caption{Fe $L_{2,3}$ (top panels) and Te $M_{4,5}$ (bottom panels) XAS (black/red curves for positive/negative helicities of the light) and XMCD (green curves) spectra for FeTe thin films with $x$ = 1 \AA/min (a,d), $x$ = 3.5 \AA/min (b, e), and $x$ = 7 \AA/min (c, f). All spectra were measured at room temperature in normal incidence. Note that the XMCD signals are multiplied with various factors that are indicated in the legends.}\label{fig_XMCD}
\end{figure*}

To characterize the electronic properties of the films, we have carried out x-ray absorption (XAS) measurements on the films across the Fe $L_{2,3}$ and Te $M_{4,5}$ edges. For comparison we have also collected the XAS spectra of a bulk single crystal with the composition Fe$_{1.14}$Te. Figure \ref{fig_XAS} shows these results. We can observe that for films with $x$ = 1 and 2 \AA/min (black curves) there are strong deviations in the Fe $L_{2,3}$ and Te $M_{4,5}$ spectra with respect to those of the Fe$_{1.14}$Te bulk reference (red curve).
For films with $x$ = 2.5, 3, 3.5, 4, and 5  \AA/min on the other hand, the spectra (black curves) are quite similar to those of the bulk (red curve). For the $x$ = 7 \AA/min film (black curve) we can, once again, observe substantial differences with the bulk (red curve). These results indicate that substantial amounts of different phases are present in the low and high $x$ films, but that for $x$ in the range of 2.5 - 5 \AA/min we are quite close to the bulk Fe$_{1+y}$Te phase.

Figure \ref{fig_XMCD} depicts Fe $L_{2,3}$ and Te $M_{4,5}$ XAS and XMCD spectra, taken at room temperature, for a representative selection of the films. For $x$ = 1 \AA/min, we observe in panel (a) that the Fe 
$L_{2,3}$ spectrum taken with the positive light helicity (black curve) is different from the 
one taken with negative light helicity (red curve). Their difference or XMCD spectrum (green 
curve) reaches a maximum value of about 18\% at 706 eV photon energy. Also at the Te $M_{4,5}$ edges there is a difference in the XAS spectra taken with the two helicities. The XMCD effect at 572 eV amounts to about 10\%. For films grown with $x$ = 3.5 and 7 \AA/min, there is essentially no difference in the XAS for the two helicities. The XMCD is smaller than 0.1-0.2 \% across both the Fe and Te edges. 

\begin{figure}[htbp]
\includegraphics[width=0.9\columnwidth]{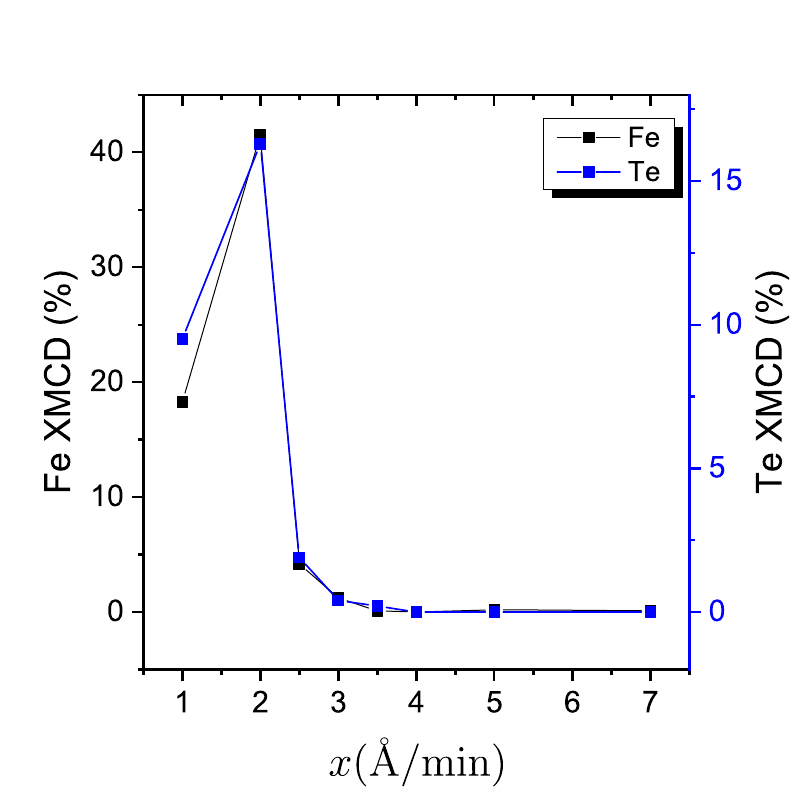}
\caption{Dependence of Fe and Te XMCD effect with Te deposition rate $x$. }\label{fig_XMCDperct}
\end{figure}

In Figure \ref{fig_XMCDperct} we summarize the XMCD results for the entire $x$ range covered in this study. We found that there is a substantial and even strong XMCD signal for films grown with $x$ = 1 and 2 \AA/min, and that this XMCD signal drops quickly for $x$ = 2.5 and 3 \AA/min films. The XMCD is tiny for $x$ = 3.5 \AA/min, and is beyond detection for the highest $x$ values, 4, 5, and 7 \AA/min. \\
From the RHEED data, we can deduce that epitaxial strain does not play a significant role, as the lattice constant is relaxed to the value expected for bulk samples. This can be due to the weakness of the van der Waals forces between layers of FeTe. Therefore, a comparison with bulk samples should be possible. For these, it is known that Fe$_{1+y}$Te with $0.06 \leq y \leq 0.16$ is paramagnetic at room temperature \cite{Koz2013}. Therefore, films that show XMCD should have Fe in excess of $y$ = 0.16. The fact that films with $x$ = 1 and 2 \AA/min have very strong XMCD may even indicate that there the excess Fe forms metallic clusters that are ferromagnetic, thereby also polarizing magnetically the adjacent Te atoms. Yet, it is also conceivable that part of the excess Fe is interstitial and then interacts with the in-plane Fe as to form magnetic clusters. Such a scenario was proposed by a study using density functional calculations \cite{Zhang2009}.

\begin{figure}[htbp]
\includegraphics[width=0.9\columnwidth]{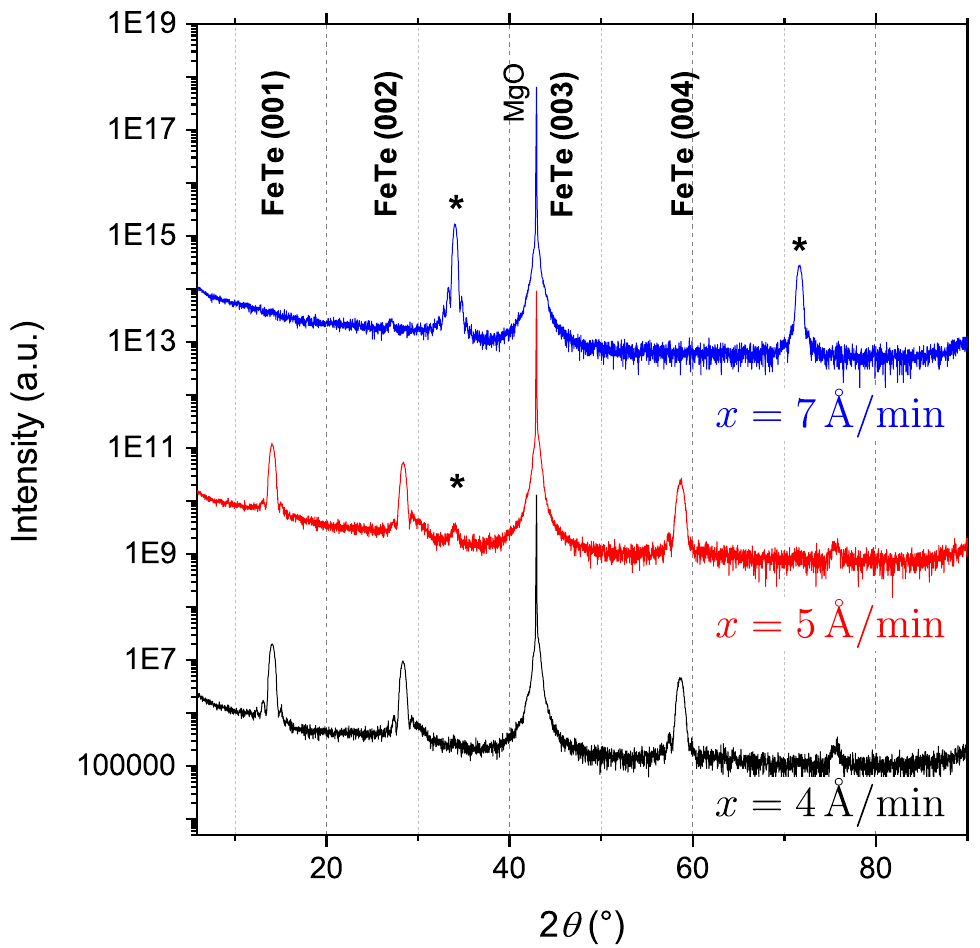}
\caption{\textit{Ex situ} XRD $\theta$ - 2$\theta$ scans of FeTe films with $x$ = 4, 5, 7 \AA/min. A new phase (marked by an asterisk) appears for $x$ = 5 \AA/min. This is the predominant phase for the film with $x$ = 7 \AA/min.}\label{fig_XRD}
\end{figure}

To complement the characterization of the films, we also have investigated \textit{ex situ} their crystal structure using XRD. A selection of XRD scans is presented in Figure \ref{fig_XRD}. For $x$ = 2.5 - 4 \AA/min, all the peaks can be indexed to FeTe (0 0 n). For $x$ = 5 \AA/min, we detect not only FeTe (0 0 n) peaks but also a secondary phase which can be identified as FeTe$_2$, marked with an asterisk in Figure \ref{fig_XRD}. This phase becomes predominant for $x$ = 7 \AA/min, with all the observed peaks belonging to it. For $x$ = 1, 2 \AA/min, no crystalline phase could be identified through XRD. 

\section{Discussion}

Our thin film experiments using the Te-limited growth reveal that epitaxial and crystalline FeTe 
films can be synthesized for a wide range of Te-deposition rates $x$. Clear and similar RHEED 
patterns can be observed for $x$ ranging from 2.5 to 5 \AA/min. Intensity oscillations of the 
specular spot in the RHEED can be recorded for films with $x$ from 2.5 to 4 \AA/min, actually 
even from $x$ = 2 \AA/min. The XRD data also show that for $x$ = 2.5 - 4 \AA/min, the films 
only show (0 0 n) peaks that belong to the FeTe phase. From the XAS measurements we find 
that, from $x$ = 2.5 to 3.5 \AA/min, the films have very similar electronic properties as the bulk 
Fe$_{1.14}$Te single crystal, and from the XMCD we see that there is no ferromagnetism for films 
with $x$ = 3.5 \AA/min or higher. From all these results combined we can infer that the 
$x$ = 3.5 \AA/min film has a composition which is very close or identical to that of the bulk 
Fe$_{1.14}$Te single crystal. This implies that the $x$ = 2.5 \AA/min film should correspond 
to a Fe$_{1.60}$Te system and that the $x$ = 4 \AA/min film should be close to a stoichiometric 
Fe$_{1.00}$Te. The latter assignment is consistent with our XRD observation that FeTe$_2$ peaks 
are visible only in films with higher $x$ values, namely $x$ = 5 and 7 \AA/min.

The picture that arises from our study is the following: films that were grown with 
$x$ = 1 and 2 \AA/min Te deposition rates have nominal compositions of roughly Fe$_4$Te and
Fe$_2$Te, respectively. They contain large amounts of presumably Fe metal clusters that
are ferromagnetic. It is nevertheless surprising that RHEED intensity oscillations can be observed
for the $x$ = 2 \AA/min film, suggesting that a substantial part of the excess Fe is residing in the interstitial positions of the FeTe phase. For $x$ = 2.5 - 3.5 \AA/min, we have a layer-by-layer growth of Fe$_{1+y}$Te films with the composition ranging from Fe$_{1.60}$Te to Fe$_{1.14}$Te, in which the excess Fe resides interstitially. For $x$ = 4 \AA/min we may have achieved the 
stoichiometric Fe$_{1.00}$Te compound and, for higher $x$ values, FeTe$_2$ starts to be formed
as well.

A note should be added regarding the purity of the Fe$_{1+y}$Te films. The presence of non-crystalline parasitic phases cannot, at this stage, be completely excluded. The electronic and sructural properties explored in this work point towards a promising route to synthesize Fe$_{1.00}$Te. However, other characterization techniques, as \textit{in situ} transport measurements, shall be also explored in the future.\\

\section{Conclusions}

To summarize, we were able to prepare good quality thin films of the Fe$_{1+y}$Te system. The 
composition ranges from Fe$_{1.00}$Te to Fe$_{1.60}$Te, which is a wider $y$ range than anticipated from studies on Fe$_{1+y}$Te bulk material \cite{Rodriguez2011, Roessler2011, Koz2013}. 
Moreover, x-ray absorption and x-ray magnetic circular dichroism measurements show that it is possible to obtain films with very similar electronic properties as that of a high quality bulk 
single crystal Fe$_{1.14}$Te. Even a high quality film with a nominally stoichiometric Fe$_{1.00}$Te composition could be made. These results suggest that molecular beam epitaxy in the Te-limited 
growth mode may provide an opportunity to synthesize FeTe in a controlled manner that can become
superconducting.

\begin{acknowledgments}
We would like to thank Sahana R{\"o}{\ss}ler and Steffen Wirth for the valuable discussions,  Katharina H{\"o}fer and Christoph Becker for the skillful technical assistance and the department of Claudia Felser for the use of the thin films XRD instrument. We acknowlegde the financial support from the Deutsche Forschungsgemeinschaft, under the Priority Program SPP-1666 Topological Insulators. C.N.W. acknowledges support from the Ministry of Science and Technology of Taiwan under MoST 105-2112-M-007-014-MY3, and V.M.P. from the International Max Planck Research School for Chemistry and Physics of Quantum Materials (IMPRS-CPQM). The experiments in Taiwan were facilitated by the Max Planck-POSTECH-Hsinchu Center for Complex Phase Materials.
\end{acknowledgments}


\begin{thebibliography}{0}%
\makeatletter
\providecommand \@ifxundefined [1]{%
 \@ifx{#1\undefined}
}%
\providecommand \@ifnum [1]{%
 \ifnum #1\expandafter \@firstoftwo
 \else \expandafter \@secondoftwo
 \fi
}%
\providecommand \@ifx [1]{%
 \ifx #1\expandafter \@firstoftwo
 \else \expandafter \@secondoftwo
 \fi
}%
\providecommand \natexlab [1]{#1}%
\providecommand \enquote  [1]{``#1''}%
\providecommand \bibnamefont  [1]{#1}%
\providecommand \bibfnamefont [1]{#1}%
\providecommand \citenamefont [1]{#1}%
\providecommand \href@noop [0]{\@secondoftwo}%
\providecommand \href [0]{\begingroup \@sanitize@url \@href}%
\providecommand \@href[1]{\@@startlink{#1}\@@href}%
\providecommand \@@href[1]{\endgroup#1\@@endlink}%
\providecommand \@sanitize@url [0]{\catcode `\\12\catcode `\$12\catcode
  `\&12\catcode `\#12\catcode `\^12\catcode `\_12\catcode `\%12\relax}%
\providecommand \@@startlink[1]{}%
\providecommand \@@endlink[0]{}%
\providecommand \url  [0]{\begingroup\@sanitize@url \@url }%
\providecommand \@url [1]{\endgroup\@href {#1}{\urlprefix }}%
\providecommand \urlprefix  [0]{URL }%
\providecommand \Eprint [0]{\href }%
\providecommand \doibase [0]{http://dx.doi.org/}%
\providecommand \selectlanguage [0]{\@gobble}%
\providecommand \bibinfo  [0]{\@secondoftwo}%
\providecommand \bibfield  [0]{\@secondoftwo}%
\providecommand \translation [1]{[#1]}%
\providecommand \BibitemOpen [0]{}%
\providecommand \bibitemStop [0]{}%
\providecommand \bibitemNoStop [0]{.\EOS\space}%
\providecommand \EOS [0]{\spacefactor3000\relax}%
\providecommand \BibitemShut  [1]{\csname bibitem#1\endcsname}%
\let\auto@bib@innerbib\@empty
\end{thebibliography}%


%


\begin{thebibliography}{25}%
\makeatletter
\providecommand \@ifxundefined [1]{%
 \@ifx{#1\undefined}
}%
\providecommand \@ifnum [1]{%
 \ifnum #1\expandafter \@firstoftwo
 \else \expandafter \@secondoftwo
 \fi
}%
\providecommand \@ifx [1]{%
 \ifx #1\expandafter \@firstoftwo
 \else \expandafter \@secondoftwo
 \fi
}%
\providecommand \natexlab [1]{#1}%
\providecommand \enquote  [1]{``#1''}%
\providecommand \bibnamefont  [1]{#1}%
\providecommand \bibfnamefont [1]{#1}%
\providecommand \citenamefont [1]{#1}%
\providecommand \href@noop [0]{\@secondoftwo}%
\providecommand \href [0]{\begingroup \@sanitize@url \@href}%
\providecommand \@href[1]{\@@startlink{#1}\@@href}%
\providecommand \@@href[1]{\endgroup#1\@@endlink}%
\providecommand \@sanitize@url [0]{\catcode `\\12\catcode `\$12\catcode
  `\&12\catcode `\#12\catcode `\^12\catcode `\_12\catcode `\%12\relax}%
\providecommand \@@startlink[1]{}%
\providecommand \@@endlink[0]{}%
\providecommand \url  [0]{\begingroup\@sanitize@url \@url }%
\providecommand \@url [1]{\endgroup\@href {#1}{\urlprefix }}%
\providecommand \urlprefix  [0]{URL }%
\providecommand \Eprint [0]{\href }%
\providecommand \doibase [0]{https://doi.org/}%
\providecommand \selectlanguage [0]{\@gobble}%
\providecommand \bibinfo  [0]{\@secondoftwo}%
\providecommand \bibfield  [0]{\@secondoftwo}%
\providecommand \translation [1]{[#1]}%
\providecommand \BibitemOpen [0]{}%
\providecommand \bibitemStop [0]{}%
\providecommand \bibitemNoStop [0]{.\EOS\space}%
\providecommand \EOS [0]{\spacefactor3000\relax}%
\providecommand \BibitemShut  [1]{\csname bibitem#1\endcsname}%
\let\auto@bib@innerbib\@empty
\bibitem [{\citenamefont {Kamihara}\ \emph {et~al.}(2008)\citenamefont
  {Kamihara}, \citenamefont {Watanabe}, \citenamefont {Hirano},\ and\
  \citenamefont {Hosono}}]{Kamihara2008}%
  \BibitemOpen
  \bibfield  {author} {\bibinfo {author} {\bibfnamefont {Y.}~\bibnamefont
  {Kamihara}}, \bibinfo {author} {\bibfnamefont {T.}~\bibnamefont {Watanabe}},
  \bibinfo {author} {\bibfnamefont {M.}~\bibnamefont {Hirano}},\ and\ \bibinfo
  {author} {\bibfnamefont {H.}~\bibnamefont {Hosono}},\ }\bibfield  {title}
  {\bibinfo {title} {Iron-based layered superconductor {La[O$_{1-x}$F$_x$]FeAs}
  ($x$= 0.05-0.12) with {T$_c$} = 26 {K}},\ }\href
  {https://doi.org/10.1021/ja800073m} {\bibfield  {journal} {\bibinfo
  {journal} {Journal of the American Chemical Society}\ }\textbf {\bibinfo
  {volume} {130}},\ \bibinfo {pages} {3296} (\bibinfo {year}
  {2008})}\BibitemShut {NoStop}%
\bibitem [{\citenamefont {Zhao}\ \emph {et~al.}(2008)\citenamefont {Zhao},
  \citenamefont {Huang}, \citenamefont {de~la Cruz}, \citenamefont {Li},
  \citenamefont {Lynn}, \citenamefont {Chen}, \citenamefont {Green},
  \citenamefont {Chen}, \citenamefont {Li}, \citenamefont {Li}, \citenamefont
  {Luo}, \citenamefont {Wang},\ and\ \citenamefont {Dai}}]{Zhao2008}%
  \BibitemOpen
  \bibfield  {author} {\bibinfo {author} {\bibfnamefont {J.}~\bibnamefont
  {Zhao}}, \bibinfo {author} {\bibfnamefont {Q.}~\bibnamefont {Huang}},
  \bibinfo {author} {\bibfnamefont {C.}~\bibnamefont {de~la Cruz}}, \bibinfo
  {author} {\bibfnamefont {S.}~\bibnamefont {Li}}, \bibinfo {author}
  {\bibfnamefont {J.~W.}\ \bibnamefont {Lynn}}, \bibinfo {author}
  {\bibfnamefont {Y.}~\bibnamefont {Chen}}, \bibinfo {author} {\bibfnamefont
  {M.~A.}\ \bibnamefont {Green}}, \bibinfo {author} {\bibfnamefont {G.~F.}\
  \bibnamefont {Chen}}, \bibinfo {author} {\bibfnamefont {G.}~\bibnamefont
  {Li}}, \bibinfo {author} {\bibfnamefont {Z.}~\bibnamefont {Li}}, \bibinfo
  {author} {\bibfnamefont {J.~L.}\ \bibnamefont {Luo}}, \bibinfo {author}
  {\bibfnamefont {N.~L.}\ \bibnamefont {Wang}},\ and\ \bibinfo {author}
  {\bibfnamefont {P.}~\bibnamefont {Dai}},\ }\bibfield  {title} {\bibinfo
  {title} {Structural and magnetic phase diagram of {CeFeAsO$_{1- x}$F$_x$} and
  its relation to high-temperature superconductivity},\ }\href
  {https://doi.org/10.1038/nmat2315} {\bibfield  {journal} {\bibinfo  {journal}
  {Nature Materials}\ }\textbf {\bibinfo {volume} {7}},\ \bibinfo {pages} {953}
  (\bibinfo {year} {2008})}\BibitemShut {NoStop}%
\bibitem [{\citenamefont {Hsu}\ \emph {et~al.}(2008)\citenamefont {Hsu},
  \citenamefont {Luo}, \citenamefont {Yeh}, \citenamefont {Chen}, \citenamefont
  {Huang}, \citenamefont {Wu}, \citenamefont {Lee}, \citenamefont {Huang},
  \citenamefont {Chu}, \citenamefont {Yan},\ and\ \citenamefont
  {Wu}}]{Hsu2008}%
  \BibitemOpen
  \bibfield  {author} {\bibinfo {author} {\bibfnamefont {F.-C.}\ \bibnamefont
  {Hsu}}, \bibinfo {author} {\bibfnamefont {J.-Y.}\ \bibnamefont {Luo}},
  \bibinfo {author} {\bibfnamefont {K.-W.}\ \bibnamefont {Yeh}}, \bibinfo
  {author} {\bibfnamefont {T.-K.}\ \bibnamefont {Chen}}, \bibinfo {author}
  {\bibfnamefont {T.-W.}\ \bibnamefont {Huang}}, \bibinfo {author}
  {\bibfnamefont {P.~M.}\ \bibnamefont {Wu}}, \bibinfo {author} {\bibfnamefont
  {Y.-C.}\ \bibnamefont {Lee}}, \bibinfo {author} {\bibfnamefont {Y.-L.}\
  \bibnamefont {Huang}}, \bibinfo {author} {\bibfnamefont {Y.-Y.}\ \bibnamefont
  {Chu}}, \bibinfo {author} {\bibfnamefont {D.-C.}\ \bibnamefont {Yan}},\ and\
  \bibinfo {author} {\bibfnamefont {M.-K.}\ \bibnamefont {Wu}},\ }\bibfield
  {title} {\bibinfo {title} {Superconductivity in the {PbO}-type structure
  $\alpha$-{FeSe}},\ }\href {https://doi.org/10.1073/pnas.0807325105}
  {\bibfield  {journal} {\bibinfo  {journal} {Proceedings of the National
  Academy of Sciences}\ }\textbf {\bibinfo {volume} {105}},\ \bibinfo {pages}
  {14262} (\bibinfo {year} {2008})}\BibitemShut {NoStop}%
\bibitem [{\citenamefont {Bao}\ \emph {et~al.}(2009)\citenamefont {Bao},
  \citenamefont {Qiu}, \citenamefont {Huang}, \citenamefont {Green},
  \citenamefont {Zajdel}, \citenamefont {Fitzsimmons}, \citenamefont
  {Zhernenkov}, \citenamefont {Chang}, \citenamefont {Fang}, \citenamefont
  {Qian}, \citenamefont {Vehstedt}, \citenamefont {Yang}, \citenamefont {Pham},
  \citenamefont {Spinu},\ and\ \citenamefont {Mao}}]{Bao2009}%
  \BibitemOpen
  \bibfield  {author} {\bibinfo {author} {\bibfnamefont {W.}~\bibnamefont
  {Bao}}, \bibinfo {author} {\bibfnamefont {Y.}~\bibnamefont {Qiu}}, \bibinfo
  {author} {\bibfnamefont {Q.}~\bibnamefont {Huang}}, \bibinfo {author}
  {\bibfnamefont {M.~A.}\ \bibnamefont {Green}}, \bibinfo {author}
  {\bibfnamefont {P.}~\bibnamefont {Zajdel}}, \bibinfo {author} {\bibfnamefont
  {M.~R.}\ \bibnamefont {Fitzsimmons}}, \bibinfo {author} {\bibfnamefont
  {M.}~\bibnamefont {Zhernenkov}}, \bibinfo {author} {\bibfnamefont
  {S.}~\bibnamefont {Chang}}, \bibinfo {author} {\bibfnamefont
  {M.}~\bibnamefont {Fang}}, \bibinfo {author} {\bibfnamefont {B.}~\bibnamefont
  {Qian}}, \bibinfo {author} {\bibfnamefont {E.~K.}\ \bibnamefont {Vehstedt}},
  \bibinfo {author} {\bibfnamefont {J.}~\bibnamefont {Yang}}, \bibinfo {author}
  {\bibfnamefont {H.~M.}\ \bibnamefont {Pham}}, \bibinfo {author}
  {\bibfnamefont {L.}~\bibnamefont {Spinu}},\ and\ \bibinfo {author}
  {\bibfnamefont {Z.~Q.}\ \bibnamefont {Mao}},\ }\bibfield  {title} {\bibinfo
  {title} {Tunable ($\delta\pi$,$\delta\pi$)-type antiferromagnetic order in
  $\alpha$-{Fe(Te,Se)} superconductors},\ }\href
  {https://doi.org/10.1103/physrevlett.102.247001} {\bibfield  {journal}
  {\bibinfo  {journal} {Physical Review Letters}\ }\textbf {\bibinfo {volume}
  {102}},\ \bibinfo {pages} {247001} (\bibinfo {year} {2009})}\BibitemShut
  {NoStop}%
\bibitem [{\citenamefont {Li}\ \emph {et~al.}(2009)\citenamefont {Li},
  \citenamefont {de~la Cruz}, \citenamefont {Huang}, \citenamefont {Chen},
  \citenamefont {Lynn}, \citenamefont {Hu}, \citenamefont {Huang},
  \citenamefont {Hsu}, \citenamefont {Yeh}, \citenamefont {Wu},\ and\
  \citenamefont {Dai}}]{Li2009}%
  \BibitemOpen
  \bibfield  {author} {\bibinfo {author} {\bibfnamefont {S.}~\bibnamefont
  {Li}}, \bibinfo {author} {\bibfnamefont {C.}~\bibnamefont {de~la Cruz}},
  \bibinfo {author} {\bibfnamefont {Q.}~\bibnamefont {Huang}}, \bibinfo
  {author} {\bibfnamefont {Y.}~\bibnamefont {Chen}}, \bibinfo {author}
  {\bibfnamefont {J.~W.}\ \bibnamefont {Lynn}}, \bibinfo {author}
  {\bibfnamefont {J.}~\bibnamefont {Hu}}, \bibinfo {author} {\bibfnamefont
  {Y.-L.}\ \bibnamefont {Huang}}, \bibinfo {author} {\bibfnamefont {F.-C.}\
  \bibnamefont {Hsu}}, \bibinfo {author} {\bibfnamefont {K.-W.}\ \bibnamefont
  {Yeh}}, \bibinfo {author} {\bibfnamefont {M.-K.}\ \bibnamefont {Wu}},\ and\
  \bibinfo {author} {\bibfnamefont {P.}~\bibnamefont {Dai}},\ }\bibfield
  {title} {\bibinfo {title} {First-order magnetic and structural phase
  transitions in {Fe$_{1+y}$SeTe$_{1-x}$}},\ }\href
  {https://doi.org/10.1103/physrevb.79.054503} {\bibfield  {journal} {\bibinfo
  {journal} {Physical Review B}\ }\textbf {\bibinfo {volume} {79}},\ \bibinfo
  {pages} {054503} (\bibinfo {year} {2009})}\BibitemShut {NoStop}%
\bibitem [{\citenamefont {Mizuguchi}\ \emph {et~al.}(2008)\citenamefont
  {Mizuguchi}, \citenamefont {Tomioka}, \citenamefont {Tsuda}, \citenamefont
  {Yamaguchi},\ and\ \citenamefont {Takano}}]{Mizuguchi2008}%
  \BibitemOpen
  \bibfield  {author} {\bibinfo {author} {\bibfnamefont {Y.}~\bibnamefont
  {Mizuguchi}}, \bibinfo {author} {\bibfnamefont {F.}~\bibnamefont {Tomioka}},
  \bibinfo {author} {\bibfnamefont {S.}~\bibnamefont {Tsuda}}, \bibinfo
  {author} {\bibfnamefont {T.}~\bibnamefont {Yamaguchi}},\ and\ \bibinfo
  {author} {\bibfnamefont {Y.}~\bibnamefont {Takano}},\ }\bibfield  {title}
  {\bibinfo {title} {Superconductivity at 27 {K} in tetragonal {FeSe} under
  high pressure},\ }\href {https://doi.org/10.1063/1.3000616} {\bibfield
  {journal} {\bibinfo  {journal} {Applied Physics Letters}\ }\textbf {\bibinfo
  {volume} {93}},\ \bibinfo {pages} {152505} (\bibinfo {year}
  {2008})}\BibitemShut {NoStop}%
\bibitem [{\citenamefont {Margadonna}\ \emph {et~al.}(2009)\citenamefont
  {Margadonna}, \citenamefont {Takabayashi}, \citenamefont {Ohishi},
  \citenamefont {Mizuguchi}, \citenamefont {Takano}, \citenamefont {Kagayama},
  \citenamefont {Nakagawa}, \citenamefont {Takata},\ and\ \citenamefont
  {Prassides}}]{Margadonna2009}%
  \BibitemOpen
  \bibfield  {author} {\bibinfo {author} {\bibfnamefont {S.}~\bibnamefont
  {Margadonna}}, \bibinfo {author} {\bibfnamefont {Y.}~\bibnamefont
  {Takabayashi}}, \bibinfo {author} {\bibfnamefont {Y.}~\bibnamefont {Ohishi}},
  \bibinfo {author} {\bibfnamefont {Y.}~\bibnamefont {Mizuguchi}}, \bibinfo
  {author} {\bibfnamefont {Y.}~\bibnamefont {Takano}}, \bibinfo {author}
  {\bibfnamefont {T.}~\bibnamefont {Kagayama}}, \bibinfo {author}
  {\bibfnamefont {T.}~\bibnamefont {Nakagawa}}, \bibinfo {author}
  {\bibfnamefont {M.}~\bibnamefont {Takata}},\ and\ \bibinfo {author}
  {\bibfnamefont {K.}~\bibnamefont {Prassides}},\ }\bibfield  {title} {\bibinfo
  {title} {Pressure evolution of the low-temperature crystal structure and
  bonding of the superconductor {FeSe} ({T$_c$}=37 {K})},\ }\href
  {https://doi.org/10.1103/physrevb.80.064506} {\bibfield  {journal} {\bibinfo
  {journal} {Physical Review B}\ }\textbf {\bibinfo {volume} {80}},\ \bibinfo
  {pages} {064506} (\bibinfo {year} {2009})}\BibitemShut {NoStop}%
\bibitem [{\citenamefont {Medvedev}\ \emph {et~al.}(2009)\citenamefont
  {Medvedev}, \citenamefont {McQueen}, \citenamefont {Troyan}, \citenamefont
  {Palasyuk}, \citenamefont {Eremets}, \citenamefont {Cava}, \citenamefont
  {Naghavi}, \citenamefont {Casper}, \citenamefont {Ksenofontov}, \citenamefont
  {Wortmann},\ and\ \citenamefont {Felser}}]{Medvedev2009}%
  \BibitemOpen
  \bibfield  {author} {\bibinfo {author} {\bibfnamefont {S.}~\bibnamefont
  {Medvedev}}, \bibinfo {author} {\bibfnamefont {T.~M.}\ \bibnamefont
  {McQueen}}, \bibinfo {author} {\bibfnamefont {I.~A.}\ \bibnamefont {Troyan}},
  \bibinfo {author} {\bibfnamefont {T.}~\bibnamefont {Palasyuk}}, \bibinfo
  {author} {\bibfnamefont {M.~I.}\ \bibnamefont {Eremets}}, \bibinfo {author}
  {\bibfnamefont {R.~J.}\ \bibnamefont {Cava}}, \bibinfo {author}
  {\bibfnamefont {S.}~\bibnamefont {Naghavi}}, \bibinfo {author} {\bibfnamefont
  {F.}~\bibnamefont {Casper}}, \bibinfo {author} {\bibfnamefont
  {V.}~\bibnamefont {Ksenofontov}}, \bibinfo {author} {\bibfnamefont
  {G.}~\bibnamefont {Wortmann}},\ and\ \bibinfo {author} {\bibfnamefont
  {C.}~\bibnamefont {Felser}},\ }\bibfield  {title} {\bibinfo {title}
  {Electronic and magnetic phase diagram of $\beta$-{Fe$_{1.01}$Se} with
  superconductivity at 36.7 {K} under pressure},\ }\href
  {https://doi.org/10.1038/nmat2491} {\bibfield  {journal} {\bibinfo  {journal}
  {Nature Materials}\ }\textbf {\bibinfo {volume} {8}},\ \bibinfo {pages} {630}
  (\bibinfo {year} {2009})}\BibitemShut {NoStop}%
\bibitem [{\citenamefont {Ciechan}\ \emph {et~al.}(2014)\citenamefont
  {Ciechan}, \citenamefont {Winiarski},\ and\ \citenamefont
  {Samsel-Czeka{\l}a}}]{Ciechan2013}%
  \BibitemOpen
  \bibfield  {author} {\bibinfo {author} {\bibfnamefont {A.}~\bibnamefont
  {Ciechan}}, \bibinfo {author} {\bibfnamefont {M.~J.}\ \bibnamefont
  {Winiarski}},\ and\ \bibinfo {author} {\bibfnamefont {M.}~\bibnamefont
  {Samsel-Czeka{\l}a}},\ }\bibfield  {title} {\bibinfo {title} {Magnetic phase
  transitions and superconductivity in strained {FeTe}},\ }\href
  {https://doi.org/10.1088/0953-8984/26/2/025702} {\bibfield  {journal}
  {\bibinfo  {journal} {Journal of Physics: Condensed Matter}\ }\textbf
  {\bibinfo {volume} {26}},\ \bibinfo {pages} {025702} (\bibinfo {year}
  {2014})}\BibitemShut {NoStop}%
\bibitem [{\citenamefont {Mydeen}\ \emph {et~al.}(2017)\citenamefont {Mydeen},
  \citenamefont {Kasinathan}, \citenamefont {Koz}, \citenamefont {R{\"o}{\ss}ler},
  \citenamefont {R{\"o}{\ss}ler}, \citenamefont {Hanfland}, \citenamefont
  {Tsirlin}, \citenamefont {Schwarz}, \citenamefont {Wirth}, \citenamefont
  {Rosner},\ and\ \citenamefont {Nicklas}}]{Mydeen2017}%
  \BibitemOpen
  \bibfield  {author} {\bibinfo {author} {\bibfnamefont {K.}~\bibnamefont
  {Mydeen}}, \bibinfo {author} {\bibfnamefont {D.}~\bibnamefont {Kasinathan}},
  \bibinfo {author} {\bibfnamefont {C.}~\bibnamefont {Koz}}, \bibinfo {author}
  {\bibfnamefont {S.}~\bibnamefont {R{\"o}{\ss}ler}}, \bibinfo {author}
  {\bibfnamefont {U.~K.}\ \bibnamefont {R{\"o}{\ss}ler}}, \bibinfo {author}
  {\bibfnamefont {M.}~\bibnamefont {Hanfland}}, \bibinfo {author}
  {\bibfnamefont {A.~A.}\ \bibnamefont {Tsirlin}}, \bibinfo {author}
  {\bibfnamefont {U.}~\bibnamefont {Schwarz}}, \bibinfo {author} {\bibfnamefont
  {S.}~\bibnamefont {Wirth}}, \bibinfo {author} {\bibfnamefont
  {H.}~\bibnamefont {Rosner}},\ and\ \bibinfo {author} {\bibfnamefont
  {M.}~\bibnamefont {Nicklas}},\ }\bibfield  {title} {\bibinfo {title}
  {Pressure-induced ferromagnetism due to an anisotropic electronic topological
  transition in {Fe$_{1.08}$Te}},\ }\href
  {https://doi.org/10.1103/physrevlett.119.227003} {\bibfield  {journal}
  {\bibinfo  {journal} {Physical Review Letters}\ }\textbf {\bibinfo {volume}
  {119}},\ \bibinfo {pages} {227003} (\bibinfo {year} {2017})}\BibitemShut
  {NoStop}%
\bibitem [{\citenamefont {Rodriguez}\ \emph {et~al.}(2011)\citenamefont
  {Rodriguez}, \citenamefont {Stock}, \citenamefont {Zajdel}, \citenamefont
  {Krycka}, \citenamefont {Majkrzak}, \citenamefont {Zavalij},\ and\
  \citenamefont {Green}}]{Rodriguez2011}%
  \BibitemOpen
  \bibfield  {author} {\bibinfo {author} {\bibfnamefont {E.~E.}\ \bibnamefont
  {Rodriguez}}, \bibinfo {author} {\bibfnamefont {C.}~\bibnamefont {Stock}},
  \bibinfo {author} {\bibfnamefont {P.}~\bibnamefont {Zajdel}}, \bibinfo
  {author} {\bibfnamefont {K.~L.}\ \bibnamefont {Krycka}}, \bibinfo {author}
  {\bibfnamefont {C.~F.}\ \bibnamefont {Majkrzak}}, \bibinfo {author}
  {\bibfnamefont {P.}~\bibnamefont {Zavalij}},\ and\ \bibinfo {author}
  {\bibfnamefont {M.~A.}\ \bibnamefont {Green}},\ }\bibfield  {title} {\bibinfo
  {title} {Magnetic-crystallographic phase diagram of the superconducting
  parent compound {Fe$_{1+x}$Te}},\ }\href
  {https://doi.org/10.1103/physrevb.84.064403} {\bibfield  {journal} {\bibinfo
  {journal} {Physical Review B}\ }\textbf {\bibinfo {volume} {84}},\ \bibinfo
  {pages} {064403} (\bibinfo {year} {2011})}\BibitemShut {NoStop}%
\bibitem [{\citenamefont {R{\"o}{\ss}ler}\ \emph {et~al.}(2011)\citenamefont
  {R{\"o}{\ss}ler}, \citenamefont {Cherian}, \citenamefont {Lorenz}, \citenamefont
  {Doerr}, \citenamefont {Koz}, \citenamefont {Curfs}, \citenamefont {Prots},
  \citenamefont {R{\"o}{\ss}ler}, \citenamefont {Schwarz}, \citenamefont
  {Elizabeth},\ and\ \citenamefont {Wirth}}]{Roessler2011}%
  \BibitemOpen
  \bibfield  {author} {\bibinfo {author} {\bibfnamefont {S.}~\bibnamefont
  {R{\"o}{\ss}ler}}, \bibinfo {author} {\bibfnamefont {D.}~\bibnamefont
  {Cherian}}, \bibinfo {author} {\bibfnamefont {W.}~\bibnamefont {Lorenz}},
  \bibinfo {author} {\bibfnamefont {M.}~\bibnamefont {Doerr}}, \bibinfo
  {author} {\bibfnamefont {C.}~\bibnamefont {Koz}}, \bibinfo {author}
  {\bibfnamefont {C.}~\bibnamefont {Curfs}}, \bibinfo {author} {\bibfnamefont
  {Y.}~\bibnamefont {Prots}}, \bibinfo {author} {\bibfnamefont {U.~K.}\
  \bibnamefont {R{\"o}{\ss}ler}}, \bibinfo {author} {\bibfnamefont
  {U.}~\bibnamefont {Schwarz}}, \bibinfo {author} {\bibfnamefont
  {S.}~\bibnamefont {Elizabeth}},\ and\ \bibinfo {author} {\bibfnamefont
  {S.}~\bibnamefont {Wirth}},\ }\bibfield  {title} {\bibinfo {title}
  {First-order structural transition in the magnetically ordered phase of
  {Fe$_{1.13}$Te}},\ }\href {https://doi.org/10.1103/physrevb.84.174506}
  {\bibfield  {journal} {\bibinfo  {journal} {Physical Review B}\ }\textbf
  {\bibinfo {volume} {84}},\ \bibinfo {pages} {174506} (\bibinfo {year}
  {2011})}\BibitemShut {NoStop}%
\bibitem [{\citenamefont {Koz}\ \emph {et~al.}(2013)\citenamefont {Koz},
  \citenamefont {R{\"o}{\ss}ler}, \citenamefont {Tsirlin}, \citenamefont {Wirth},\
  and\ \citenamefont {Schwarz}}]{Koz2013}%
  \BibitemOpen
  \bibfield  {author} {\bibinfo {author} {\bibfnamefont {C.}~\bibnamefont
  {Koz}}, \bibinfo {author} {\bibfnamefont {S.}~\bibnamefont {R{\"o}{\ss}ler}},
  \bibinfo {author} {\bibfnamefont {A.~A.}\ \bibnamefont {Tsirlin}}, \bibinfo
  {author} {\bibfnamefont {S.}~\bibnamefont {Wirth}},\ and\ \bibinfo {author}
  {\bibfnamefont {U.}~\bibnamefont {Schwarz}},\ }\bibfield  {title} {\bibinfo
  {title} {Low-temperature phase diagram of {Fe$_{1+y}$Te} studied using x-ray
  diffraction},\ }\href {https://doi.org/10.1103/physrevb.88.094509} {\bibfield
   {journal} {\bibinfo  {journal} {Physical Review B}\ }\textbf {\bibinfo
  {volume} {88}},\ \bibinfo {pages} {094509} (\bibinfo {year}
  {2013})}\BibitemShut {NoStop}%
\bibitem [{\citenamefont {Han}\ \emph {et~al.}(2010)\citenamefont {Han},
  \citenamefont {Li}, \citenamefont {Cao}, \citenamefont {Wang}, \citenamefont
  {Xu}, \citenamefont {Zhao}, \citenamefont {Guo},\ and\ \citenamefont
  {Yang}}]{Han2010}%
  \BibitemOpen
  \bibfield  {author} {\bibinfo {author} {\bibfnamefont {Y.}~\bibnamefont
  {Han}}, \bibinfo {author} {\bibfnamefont {W.~Y.}\ \bibnamefont {Li}},
  \bibinfo {author} {\bibfnamefont {L.~X.}\ \bibnamefont {Cao}}, \bibinfo
  {author} {\bibfnamefont {X.~Y.}\ \bibnamefont {Wang}}, \bibinfo {author}
  {\bibfnamefont {B.}~\bibnamefont {Xu}}, \bibinfo {author} {\bibfnamefont
  {B.~R.}\ \bibnamefont {Zhao}}, \bibinfo {author} {\bibfnamefont {Y.~Q.}\
  \bibnamefont {Guo}},\ and\ \bibinfo {author} {\bibfnamefont {J.~L.}\
  \bibnamefont {Yang}},\ }\bibfield  {title} {\bibinfo {title}
  {Superconductivity in iron telluride thin films under tensile stress},\
  }\href {https://doi.org/10.1103/physrevlett.104.017003} {\bibfield  {journal}
  {\bibinfo  {journal} {Physical Review Letters}\ }\textbf {\bibinfo {volume}
  {104}},\ \bibinfo {pages} {017003} (\bibinfo {year} {2010})}\BibitemShut
  {NoStop}%
\bibitem [{\citenamefont {Nie}\ \emph {et~al.}(2010)\citenamefont {Nie},
  \citenamefont {Telesca}, \citenamefont {Budnick}, \citenamefont {Sinkovic},\
  and\ \citenamefont {Wells}}]{Nie2010}%
  \BibitemOpen
  \bibfield  {author} {\bibinfo {author} {\bibfnamefont {Y.~F.}\ \bibnamefont
  {Nie}}, \bibinfo {author} {\bibfnamefont {D.}~\bibnamefont {Telesca}},
  \bibinfo {author} {\bibfnamefont {J.~I.}\ \bibnamefont {Budnick}}, \bibinfo
  {author} {\bibfnamefont {B.}~\bibnamefont {Sinkovic}},\ and\ \bibinfo
  {author} {\bibfnamefont {B.~O.}\ \bibnamefont {Wells}},\ }\bibfield  {title}
  {\bibinfo {title} {Superconductivity induced in iron telluride films by
  low-temperature oxygen incorporation},\ }\href
  {https://doi.org/10.1103/physrevb.82.020508} {\bibfield  {journal} {\bibinfo
  {journal} {Physical Review B}\ }\textbf {\bibinfo {volume} {82}},\ \bibinfo
  {pages} {020508(R)} (\bibinfo {year} {2010})}\BibitemShut {NoStop}%
\bibitem [{\citenamefont {Si}\ \emph {et~al.}(2010)\citenamefont {Si},
  \citenamefont {Jie}, \citenamefont {Wu}, \citenamefont {Zhou}, \citenamefont
  {Gu}, \citenamefont {Johnson},\ and\ \citenamefont {Li}}]{Si2010}%
  \BibitemOpen
  \bibfield  {author} {\bibinfo {author} {\bibfnamefont {W.}~\bibnamefont
  {Si}}, \bibinfo {author} {\bibfnamefont {Q.}~\bibnamefont {Jie}}, \bibinfo
  {author} {\bibfnamefont {L.}~\bibnamefont {Wu}}, \bibinfo {author}
  {\bibfnamefont {J.}~\bibnamefont {Zhou}}, \bibinfo {author} {\bibfnamefont
  {G.}~\bibnamefont {Gu}}, \bibinfo {author} {\bibfnamefont {P.~D.}\
  \bibnamefont {Johnson}},\ and\ \bibinfo {author} {\bibfnamefont
  {Q.}~\bibnamefont {Li}},\ }\bibfield  {title} {\bibinfo {title}
  {Superconductivity in epitaxial thin films of {Fe$_{1.08}$Te:O$_x$}},\ }\href
  {https://doi.org/10.1103/physrevb.81.092506} {\bibfield  {journal} {\bibinfo
  {journal} {Physical Review B}\ }\textbf {\bibinfo {volume} {81}},\ \bibinfo
  {pages} {092506} (\bibinfo {year} {2010})}\BibitemShut {NoStop}%
\bibitem [{\citenamefont {Telesca}\ \emph {et~al.}(2012)\citenamefont
  {Telesca}, \citenamefont {Nie}, \citenamefont {Budnick}, \citenamefont
  {Wells},\ and\ \citenamefont {Sinkovic}}]{Telesca2012}%
  \BibitemOpen
  \bibfield  {author} {\bibinfo {author} {\bibfnamefont {D.}~\bibnamefont
  {Telesca}}, \bibinfo {author} {\bibfnamefont {Y.}~\bibnamefont {Nie}},
  \bibinfo {author} {\bibfnamefont {J.~I.}\ \bibnamefont {Budnick}}, \bibinfo
  {author} {\bibfnamefont {B.~O.}\ \bibnamefont {Wells}},\ and\ \bibinfo
  {author} {\bibfnamefont {B.}~\bibnamefont {Sinkovic}},\ }\bibfield  {title}
  {\bibinfo {title} {Impact of valence states on the superconductivity of iron
  telluride and iron selenide films with incorporated oxygen},\ }\href
  {https://doi.org/10.1103/physrevb.85.214517} {\bibfield  {journal} {\bibinfo
  {journal} {Physical Review B}\ }\textbf {\bibinfo {volume} {85}},\ \bibinfo
  {pages} {214517} (\bibinfo {year} {2012})}\BibitemShut {NoStop}%
\bibitem [{\citenamefont {Zheng}(2013)}]{Zheng2013}%
  \BibitemOpen
  \bibfield  {author} {\bibinfo {author} {\bibfnamefont {M.}~\bibnamefont
  {Zheng}},\ }\emph {\bibinfo {title} {Superconductivity in oxygen doped iron
  telluride by molecular beam epitaxy}},\ \href@noop {} {Ph.D. thesis},\
  \bibinfo  {school} {University of Illinois at Urbana-Champaign} (\bibinfo
  {year} {2013})\BibitemShut {NoStop}%
\bibitem [{\citenamefont {Dong}\ \emph {et~al.}(2011)\citenamefont {Dong},
  \citenamefont {Wang}, \citenamefont {Li}, \citenamefont {Chen}, \citenamefont
  {Yuan},\ and\ \citenamefont {Fang}}]{Dong2011}%
  \BibitemOpen
  \bibfield  {author} {\bibinfo {author} {\bibfnamefont {C.}~\bibnamefont
  {Dong}}, \bibinfo {author} {\bibfnamefont {H.}~\bibnamefont {Wang}}, \bibinfo
  {author} {\bibfnamefont {Z.}~\bibnamefont {Li}}, \bibinfo {author}
  {\bibfnamefont {J.}~\bibnamefont {Chen}}, \bibinfo {author} {\bibfnamefont
  {H.~Q.}\ \bibnamefont {Yuan}},\ and\ \bibinfo {author} {\bibfnamefont
  {M.}~\bibnamefont {Fang}},\ }\bibfield  {title} {\bibinfo {title} {Revised
  phase diagram for the {FeTe$_{1-x}$Se$_x$} system with fewer excess {Fe}
  atoms},\ }\href {https://doi.org/10.1103/physrevb.84.224506} {\bibfield
  {journal} {\bibinfo  {journal} {Physical Review B}\ }\textbf {\bibinfo
  {volume} {84}},\ \bibinfo {pages} {224506} (\bibinfo {year}
  {2011})}\BibitemShut {NoStop}%
\bibitem [{\citenamefont {Hu}\ \emph {et~al.}(2012)\citenamefont {Hu},
  \citenamefont {Wang}, \citenamefont {Qian},\ and\ \citenamefont
  {Mao}}]{Hu2012}%
  \BibitemOpen
  \bibfield  {author} {\bibinfo {author} {\bibfnamefont {J.}~\bibnamefont
  {Hu}}, \bibinfo {author} {\bibfnamefont {G.~C.}\ \bibnamefont {Wang}},
  \bibinfo {author} {\bibfnamefont {B.}~\bibnamefont {Qian}},\ and\ \bibinfo
  {author} {\bibfnamefont {Z.~Q.}\ \bibnamefont {Mao}},\ }\bibfield  {title}
  {\bibinfo {title} {Inhomogeneous superconductivity induced by interstitial
  {Fe} deintercalation in oxidizing-agent-annealed and {HNO$_3$}-treated
  {Fe$_{1+y}$(Te$_{1-x}$Se$_x$)}},\ }\href
  {https://doi.org/10.1088/0953-2048/25/8/084011} {\bibfield  {journal}
  {\bibinfo  {journal} {Superconductor Science and Technology}\ }\textbf
  {\bibinfo {volume} {25}},\ \bibinfo {pages} {084011} (\bibinfo {year}
  {2012})}\BibitemShut {NoStop}%
\bibitem [{\citenamefont {Kawasaki}\ \emph {et~al.}(2012)\citenamefont
  {Kawasaki}, \citenamefont {Deguchi}, \citenamefont {Demura}, \citenamefont
  {Watanabe}, \citenamefont {Okazaki}, \citenamefont {Ozaki}, \citenamefont
  {Yamaguchi}, \citenamefont {Takeya},\ and\ \citenamefont
  {Takano}}]{Kawasaki2012}%
  \BibitemOpen
  \bibfield  {author} {\bibinfo {author} {\bibfnamefont {Y.}~\bibnamefont
  {Kawasaki}}, \bibinfo {author} {\bibfnamefont {K.}~\bibnamefont {Deguchi}},
  \bibinfo {author} {\bibfnamefont {S.}~\bibnamefont {Demura}}, \bibinfo
  {author} {\bibfnamefont {T.}~\bibnamefont {Watanabe}}, \bibinfo {author}
  {\bibfnamefont {H.}~\bibnamefont {Okazaki}}, \bibinfo {author} {\bibfnamefont
  {T.}~\bibnamefont {Ozaki}}, \bibinfo {author} {\bibfnamefont
  {T.}~\bibnamefont {Yamaguchi}}, \bibinfo {author} {\bibfnamefont
  {H.}~\bibnamefont {Takeya}},\ and\ \bibinfo {author} {\bibfnamefont
  {Y.}~\bibnamefont {Takano}},\ }\bibfield  {title} {\bibinfo {title} {Phase
  diagram and oxygen annealing effect of {FeTe$_{1-x}$Se$_x$} iron-based
  superconductor},\ }\href {https://doi.org/10.1016/j.ssc.2012.04.002}
  {\bibfield  {journal} {\bibinfo  {journal} {Solid State Communications}\
  }\textbf {\bibinfo {volume} {152}},\ \bibinfo {pages} {1135} (\bibinfo {year}
  {2012})}\BibitemShut {NoStop}%
\bibitem [{\citenamefont {Hu}\ \emph {et~al.}(2014)\citenamefont {Hu},
  \citenamefont {Kwon}, \citenamefont {Zheng}, \citenamefont {Zhang},
  \citenamefont {Greene}, \citenamefont {Eckstein},\ and\ \citenamefont
  {Zuo}}]{Hu2014}%
  \BibitemOpen
  \bibfield  {author} {\bibinfo {author} {\bibfnamefont {H.}~\bibnamefont
  {Hu}}, \bibinfo {author} {\bibfnamefont {J.-H.}\ \bibnamefont {Kwon}},
  \bibinfo {author} {\bibfnamefont {M.}~\bibnamefont {Zheng}}, \bibinfo
  {author} {\bibfnamefont {C.}~\bibnamefont {Zhang}}, \bibinfo {author}
  {\bibfnamefont {L.~H.}\ \bibnamefont {Greene}}, \bibinfo {author}
  {\bibfnamefont {J.~N.}\ \bibnamefont {Eckstein}},\ and\ \bibinfo {author}
  {\bibfnamefont {J.-M.}\ \bibnamefont {Zuo}},\ }\bibfield  {title} {\bibinfo
  {title} {Impact of interstitial oxygen on the electronic and magnetic
  structure in superconducting {Fe$_{1+y}$TeO$_x$} thin films},\ }\href
  {https://doi.org/10.1103/physrevb.90.180504} {\bibfield  {journal} {\bibinfo
  {journal} {Physical Review B}\ }\textbf {\bibinfo {volume} {90}},\ \bibinfo
  {pages} {180504(R)} (\bibinfo {year} {2014})}\BibitemShut {NoStop}%
\bibitem [{\citenamefont {Li}\ \emph {et~al.}(2016)\citenamefont {Li},
  \citenamefont {Yin}, \citenamefont {Wang}, \citenamefont {He}, \citenamefont
  {Ma}, \citenamefont {Xue},\ and\ \citenamefont {Chen}}]{Li2016}%
  \BibitemOpen
  \bibfield  {author} {\bibinfo {author} {\bibfnamefont {W.}~\bibnamefont
  {Li}}, \bibinfo {author} {\bibfnamefont {W.-G.}\ \bibnamefont {Yin}},
  \bibinfo {author} {\bibfnamefont {L.}~\bibnamefont {Wang}}, \bibinfo {author}
  {\bibfnamefont {K.}~\bibnamefont {He}}, \bibinfo {author} {\bibfnamefont
  {X.}~\bibnamefont {Ma}}, \bibinfo {author} {\bibfnamefont {Q.-K.}\
  \bibnamefont {Xue}},\ and\ \bibinfo {author} {\bibfnamefont {X.}~\bibnamefont
  {Chen}},\ }\bibfield  {title} {\bibinfo {title} {Charge ordering in
  stoichiometric {FeTe}: Scanning tunneling microscopy and spectroscopy},\
  }\href {https://doi.org/10.1103/physrevb.93.041101} {\bibfield  {journal}
  {\bibinfo  {journal} {Physical Review B}\ }\textbf {\bibinfo {volume} {93}},\
  \bibinfo {pages} {041101(R)} (\bibinfo {year} {2016})}\BibitemShut {NoStop}%
\bibitem [{\citenamefont {Rö{\ss}ler}\ \emph {et~al.}(2016)\citenamefont
  {Rö{\ss}ler}, \citenamefont {Koz}, \citenamefont {Wirth},\ and\
  \citenamefont {Schwarz}}]{Roesler2016}%
  \BibitemOpen
  \bibfield  {author} {\bibinfo {author} {\bibfnamefont {S.}~\bibnamefont
  {Rö{\ss}ler}}, \bibinfo {author} {\bibfnamefont {C.}~\bibnamefont {Koz}},
  \bibinfo {author} {\bibfnamefont {S.}~\bibnamefont {Wirth}},\ and\ \bibinfo
  {author} {\bibfnamefont {U.}~\bibnamefont {Schwarz}},\ }\bibfield  {title}
  {\bibinfo {title} {Synthesis, phase stability, structural, and physical
  properties of 11-type iron chalcogenides},\ }\href
  {https://doi.org/10.1002/pssb.201600149} {\bibfield  {journal} {\bibinfo
  {journal} {Physica Status Solidi (b)}\ }\textbf {\bibinfo {volume} {254}},\
  \bibinfo {pages} {1600149} (\bibinfo {year} {2016})}\BibitemShut {NoStop}%
\bibitem [{\citenamefont {Zhang}\ \emph {et~al.}(2009)\citenamefont {Zhang},
  \citenamefont {Singh},\ and\ \citenamefont {Du}}]{Zhang2009}%
  \BibitemOpen
  \bibfield  {author} {\bibinfo {author} {\bibfnamefont {L.}~\bibnamefont
  {Zhang}}, \bibinfo {author} {\bibfnamefont {D.~J.}\ \bibnamefont {Singh}},\
  and\ \bibinfo {author} {\bibfnamefont {M.~H.}\ \bibnamefont {Du}},\
  }\bibfield  {title} {\bibinfo {title} {Density functional study of excess
  {Fe} in {Fe$_{1+x}$Te}: Magnetism and doping},\ }\href
  {https://doi.org/10.1103/physrevb.79.012506} {\bibfield  {journal} {\bibinfo
  {journal} {Physical Review B}\ }\textbf {\bibinfo {volume} {79}},\ \bibinfo
  {pages} {012506} (\bibinfo {year} {2009})}\BibitemShut {NoStop}%
\end{thebibliography}
\end{document}